\documentclass[nofootinbib,aps,pra,twocolumn,secnumarabic,balancelastpage,amsmath,amssymb,floatfix,superscriptaddress]{revtex4-1}

\usepackage{xcolor}

\usepackage{bm}
\usepackage{ulem}
\usepackage{mathtools}
\newcommand{\tr}{\operatorname{Tr}}
\newcommand{\ketbra}[2]{\ket {#1} \hskip -0.8ex \bra {#2}}

\newcommand{\expcth}[1]{\left\langle #1 \right\rangle_{\mathrm{Haar}} }
\newcommand{\Gammalr}[1]{\Gamma\left(#1\right)}

\newcommand{\Renyi}{R\'enyi }

\newcommand{\characteristic}{ \expcth{e^{z|W(p,q)|}}  }

\usepackage{amssymb}
\usepackage{amsmath}
\usepackage{amsfonts}
\usepackage{braket}
\usepackage{graphicx}
\usepackage{epstopdf}
\usepackage{epsfig}
\usepackage[toc,page,title,titletoc,header]{appendix}
\usepackage{dsfont,amsthm,amsbsy}
\usepackage{hyperref}
\usepackage[xspace]{ellipsis}
\usepackage{cancel}

\DeclareMathOperator{\erf}{erf}

\begin{document}
\title{Mana in Haar-random states}
\author{Justin H. Wilson}
\affiliation{Department of Physics and Astronomy, Center for Materials Theory, Rutgers University, Piscataway, NJ 08854 USA}

\author{Christopher David White}
\affiliation{Condensed Matter Theory Center, University of Maryland, College Park, Md, 20742}
\email{cdwhite@umd.edu}

\begin{abstract}
  Mana is a measure of the amount of non-Clifford resources required to create a state;
  the mana of a mixed state on $\ell$ qudits bounded by
  $\le \frac 1 2 (\ell \ln d - S_2)$;
  $S_2$ the state's second \Renyi entropy.
  We compute the mana of Haar-random pure and mixed states and
  find that the mana is nearly logarithmic in Hilbert space dimension:
  that is, extensive in number of qudits and logarithmic in qudit dimension.
  In particular, the average mana of states with less-than-maximal entropy falls short of that maximum by $\ln \pi/2$.
  We then connect this result to recent work on near-Clifford approximate $t$-designs;
  in doing so we point out that mana is a useful measure of non-Clifford resources precisely because it is not differentiable.
\end{abstract}

\maketitle

\section{Introduction}

The emergence of thermalization from unitary quantum dynamics has been a focus of recent many-body physics.
Local random unitary dynamics give a ``minimally structured'' model of thermalization~\cite{nahum_quantum_2017}
that lacks even energy conservation.
These models clarify the links between information scrambling, thermalization, and the dynamics of black holes~\cite{hayden_black_2007}.
One can further introduce measurements and observe a so-called ``meaurement-induced phase transition'' in the entanglement;
as the system is measured more frequently, the entanglement transitions from area-law to volume-law~\cite{li_quantum_2018,skinner_measurement_2019,vasseur_entanglement_2019}.
In the study of quantum computation,
random unitaries also inspired certain methods for characterizing quantum gates:
since gate errors in a string of random unitaries average out to a depolarizing channel,
schemes involving random circuits---called ``randomized benchmarking''---can replace full quantum process tomography~\cite{emerson_scalable_2005}.
% This scrambling occurs even without globally conserved energy (such as in the case of the maximal scrambler, a black hole~\cite{hayden_black_2007}); leading to the study of ``minimally structured'' random unitary models~\cite{nahum_quantum_2017}.
% Random unitaries model thermalization and the dynamics of black holes \cite{hayden_black_2007},
% and play a key role in ``minimally structured'' models of local thermalization in a condensed matter context \cite{nahum_quantum_2017}.
% From these minimally structured models grew the wide world of measurement-induced phase transitions.\todo{Justin please do something about this terrible sentence}

But Haar-random unitaries are not easy to implement,
either on a classical computer or an error-corrected quantum computer.
No quantum error correcting code can support transversal implementations of every gate in a universal quantum gate set~\cite{eastin_restrictions_2009}.
In most codes Clifford gates are transversal,
and one imagines implementing non-Clifford gates 
by magic-state injection schemes
requiring a costly magic-state distillation procedure~\cite{bravyi_universal_2005}.
Non-Clifford gates will therefore control the difficulty of error-corrected quantum computations.
%in much the same way that entangling gates control the difficulty of quantum computations on NISQ \cite{todo} devices \cite{todo}.
At the same time, classical computers can efficiently simulate circuits consisting mostly or entirely of Clifford gates~\cite{aaronson_improved_2004,anders_fast_2006,bravyi_trading_2016,bravyi_improved_2016,howard_application_2017,bennink_unbiased_2017,bravyi_simulation_2019,bu_efficient_2019,huang_approximate_2019}.
% The non-Clifford operations therefore control the difficulty of classical simulations
% in much the same way that entangling gates control the difficulty of matrix product state simulations.

The importance of random unitaries
and the relative ease of implementing Clifford gates
has prompted interest in Clifford or near-Clifford models of random unitaries.
Such a model is called a $t$-design:
$t$-designs are distribution that matches first $t$ moments of the Haar distribution~\cite{dankert_exact_2009}.
A 2-design, therefore, suffices for randomized benchmarking.
The Clifford group is a 2-design,
and has formed the basis for many applications of randomized benchmarking~\cite{knill_randomized_2008,magesan_scalable_2011,magesan_scalable_2011,magesan_characterizing_2012,magesan_characterizing_2012,gambetta_characterization_2012,morvan_qutrit_2020}.
In fact, random Clifford circuits on qubits are known to form 3- but not 4-designs~\cite{webb_clifford_2016,zhu_multiqubit_2017};
the addition of a handful of non-Clifford gates results can give an approximate $t$-design for any $t$~\cite{haferkamp_quantum_2020}.

\begin{figure}[t]
  \begin{minipage}{0.45\textwidth}
    \includegraphics[width=\textwidth]{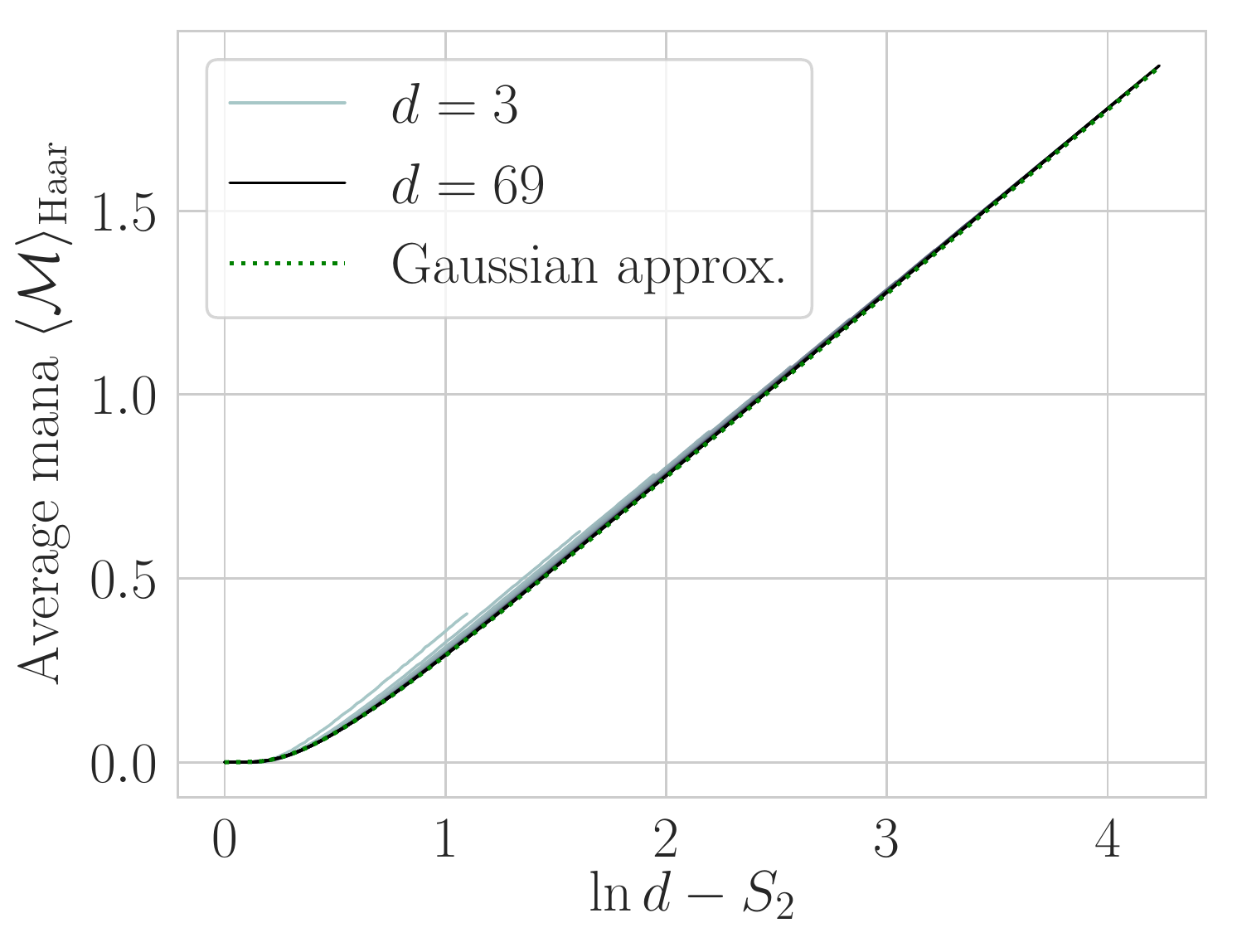}
  \end{minipage}
  
  \caption{
    \textbf{Random mixed state mana}: average mana of numerically sampled random mixed states on a single qudit of dimension $d$
    as a function of entropy deficit $\Delta = \ln d - S_2$,
    together with the asymptotic Gaussian approximation of \eqref{eq:gaussian-mixed} with \eqref{eq:sigma-mu-Delta} (green dotted line)
    for some odd Hilbert space dimensions from $d = 3$ to $d = 69$.
    We average over 4000 mixed states at each entropy.
    For $d \gtrsim 5$ (second grey line from top)
    the numerical average is very close to the asymptotic Gaussian prediction,
    and for $d = 69$ it is indistinguishable on this plot
    (but see Fig.~\ref{fig:mixed-detail}, where finite-size effects are more apparent).
  }
  \label{fig:mixed-intro}
\end{figure}

This raises the question:
If we use a near-Clifford construction like the approximate $t$-designs of~\cite{haferkamp_quantum_2020} to model some physical process like thermalization, what physics will we miss?
The measurement-induced phase transitions, for example, behaves differently when the random unitaries involved are Clifford, as opposed to Haar~\cite{zabalo_critical_2020}.
Additionally, Clifford gates constitute classical dynamics
in the sense that efficient classical algorithms exist for the dynamics.
So understanding the degree to which thermalization can be modeled by Clifford dynamics
may clarify the relationship
between quantum thermalization,
which is understood in terms of the eigenstate thermalization hypothesis~\cite{deutsch_quantum_1991,srednicki_chaos_1994},
and classical thermalization,
which can be understood via dynamical chaos~\cite{gallavotti_statistical_2013},
and provide a practical route to more efficient algorithms.
And estimates of the non-Clifford resources required to prepare a state
will constrain tensor network models of that state
(cf \cite{white_conformal_2020}, which used such considerations to constrain the MERAs that construct CFT ground states).

We study the degree to which models of random unitaries require non-Clifford resources
by computing a magic monotone called ``mana'' on Haar-random states.
Magic monotones functions on quantum states
that are non-increasing under certain operations,
roughly the Clifford gates \cite{veitch_resource_2014}.
They therefore measure the amount of non-Clifford resources rquired to create the state.
Many such functions exist \cite{howard_application_2016,wang_efficiently_2018,regula_convex_2018,bravyi_simulation_2019,beverland_lower_2019}.
We choose the mana of \cite{veitch_resource_2014} for its tractability.
Mana is related to the negativity of the discrete Wigner function;
it is defined on (systems of) qudits of odd dimension.
The mana of a state $\ket \psi$ gives lower bounds on both
the number of copies of $\ket \psi$ required to distill a magic state suitable for injection,
and the number of magic states required to prepare $\ket \psi$.
A (generically mixed) state on $\ell$ qudits of dimension $d$ has mana bounded by
\begin{equation}\label{eq:intro-jensen}
  \mathcal M \le \frac 1 2[\ell \ln d - S_2]
\end{equation}
where $S_2$ is the state's second \Renyi entropy, by Jensen's inequality.

We compute the mana for pure and mixed states drawn at random from distributions invariant under unitaries.
We find that it is
the maximum \eqref{eq:intro-jensen}
modified by a constant correction $\frac 1 2 \ln \pi/2$ (cf.\ Fig.~\ref{fig:mixed-intro}).
For $d \gg 1$ almost all states have mana very close to this value.
Consequently mana distinguishes Haar states from the output of near-Clifford $t$-designs;
this is a consequence of the fact that all moments are relevant when calculating the mana
(when considered as a function of wavefunction components, it has a cusp).

We proceed as follows.
In Sec.~\ref{s:bg} we review basic definitions,
and in Sec.~\ref{s:summary} we summarize our results.
In Sec.~\ref{s:mixed} we compute the mana for random mixed states.
We offer first a heuristic Gaussian calculation,
which is straightforward, illuminating, and asymptotically correct;
and then a detailed exact calculation.
In Sec.~\ref{s:pure} we consider the pure state special case in more detail,
and in Sec.~\ref{s:tdesign} we argue that the mana distinguishes Haar states from near-Clifford approximate $t$-designs,
and use Fourier and Chebyshev expansions to explain this fact in terms of mana's cusp.
In Sec.~\ref{s:discussion} we discuss some implications of our results.

\section{Mana and the Haar distribution}\label{s:bg}

\subsection{Mana: definitions}\label{s:mana-def}
Consider first a state $\ket \psi = \sum_j \psi_j \ket j$ on a Hilbert space of odd dimension $d$.
The \textbf{discrete Wigner function} of $\ket \psi$ is the discrete analog of the continuum Wigner function:
\begin{equation}
  \label{eq:wigner-def}
  W(p,q) = d^{-1} \sum_{x} \omega^{-px} \psi_{q + 2^{-1} x} \psi^*_{q -2^{-1} x}\;;
\end{equation}
where $\omega = e^{2\pi i / d}$ and $2^{-1} \equiv (d+1)/2$ is the inverse of $2$ in the field $\mathbb Z_d$.
We can write the discrete Wigner function as a set of operator expectation values
\begin{equation}
  \label{eq:wigner-defn-expct}
  W(p,q) = d^{-1} \bra \psi A(p,q) \ket \psi\;,
\end{equation}
where the operators $A(p,q)$,
called \textbf{phase space point operators}.
These operators
are Hermitian and unitarily equivalent with spectrum
\begin{equation}
  \label{eq:A-spec}
  \mathrm{spec}\;A(p,q) =
  [
  \underbrace{1, \dots, 1}_{\text{$(d + 1)/2$} },
  \underbrace{-1, \dots, -1}_{\text{$(d - 1)/2$} }
  ]
\end{equation}
and trace inner product
\begin{equation}
  \label{eq:A-tr-ip}
  \tr[A(p,q) A(p',q')] = d \delta_{pp'}\delta_{qq'}\;.
\end{equation}
With $d^{2}$ operators $A(p,q)$, \eqref{eq:A-tr-ip} also implies that these are a complete set of operators.
(We discuss the phase space point operators in more detail in App.~\ref{app:pauli-phase}.)
The discrete Wigner function of a mixed state $\rho$ is the natural generalization of \eqref{eq:wigner-defn-expct}
\begin{equation}
  W(p,q) = d^{-1} \tr[ \rho A(p,q)]\;.
\end{equation}
The Wigner function of a state is constrained by
\begin{subequations}
  \label{eq:wigner-norm-trace-constraint}
  \begin{align}
    \sum_{p,q} W(p,q) &= 1\\
    \sum_{p,q} W(p,q)^2 &= \frac 1 d e^{-S_2} \label{eq:wigner-sum-sq}\;, 
  \end{align}
\end{subequations}
$S_2$ the state's second \Renyi entropy;
one can see this using the spectrum \eqref{eq:A-spec}, which gives $\tr A(p,q) = 1$,
and the inner product \eqref{eq:A-tr-ip}.

If $\ket \psi$ is a stabilizer state (an eigenstate of a generalized Pauli operator, cf.\ App.~\ref{app:pauli-phase})
or a statistical mixture of stabilizer states,
then its Wigner function is non-negative \cite{gross_hudsons_2006,gross_non-negative_2007}.
The converse is true only for pure states:
every pure state with non-negative Wigner function is a stabilizer state,
but not all mixed states with non-negative Wigner function can be constructed as statistical mixtures of stabilizer states
(see \cite[Sec.~V]{gross_hudsons_2006} for a concrete example and \cite{veitch_negative_2012} for a geometrical discussion).

The \textbf{Wigner norm} measures the degree to which a state's Wigner function is negative.
The Wigner norm of a density matrix $\rho$ is the $L^1$ norm in a basis of normalized phase-space point operators
\begin{equation}
  \mathcal W \equiv \| \rho \|_W := \sum_{p,q} |W(p,q)|\;;
\end{equation}
the \textbf{mana} is
\begin{equation}
  \mathcal M = \ln \mathcal W\;.
\end{equation}
Applying Jensen's inequality to the Wigner norm and using \eqref{eq:wigner-sum-sq},
we find
\begin{equation}
  \label{eq:jensens-bound}
  \mathcal M \le \frac 1 2 [\ln d - S_2]\;.
\end{equation}

\subsection{Wigner functions and mana on multi-qubit systems}
In Sec.~\ref{s:mana-def} we defined the phase space point operators, Wigner function, Wigner norm, and mana
for a single dimension-$d$ qudit.
We can extend these definitions to multi-qudit systems in either of two ways:
a trivial ``single-particle'' definition
in which we understand an $\ell$-qudit system as having an undifferentiated $d^\ell$-dimensional Hilbert space
and stroll blithely through the definitions of \ref{s:mana-def},
and a richer ``many-particle'' definition.
The calculations in this work proceed identically in the two extensions.

In the ``many-particle'' extension of \ref{s:mana-def},  
we define $\ell$-qudit phase space point operators
\begin{equation}
  \label{eq:ell-psp}
  A(\bm p, \bm q) = A(p_1,q_1) \otimes \dots \otimes A(p_\ell, q_\ell)
\end{equation}
and replace the Hilbert space dimension $d$ by $d^\ell$ where necessary.
The Wigner function is then
\begin{equation}
  W(\bm p, \bm q) = d^{-\ell} \tr[\rho A(\bm p, \bm q)]
\end{equation}
and the Wigner norm and mana as before
\begin{equation}
  \mathcal W = \sum_{\bm p \bm q} |W(\bm p, \bm q)|,\qquad \mathcal M = \ln \mathcal W.
\end{equation}
The Jensen's inequality bound is 
\begin{equation}
  \mathcal M \le \frac 1 2 [\ell \ln d - S_2]\;.
\end{equation}
The Wigner norm and the mana are ``magic monotones'',
meaning that they are nonincreasing under Clifford operations,
partial traces,
and measurements of Pauli operators \cite{veitch_resource_2014}.
They therefore measure the amount of non-Clifford resources required to create a state.

The $\ell$-qudit phase space point operators of Eq.~\ref{eq:ell-psp}
are not identical to the na\"ive ``single-particle'' phase space point operators of Sec.~\ref{s:mana-def}
taken on $d^\ell$ dimensions,
but they have the same spectrum \eqref{eq:A-spec} (with $d \mapsto d^\ell$).
Since our calculations below depend solely on this spectrum,
we need not concern ourselves with the distinction between ``single-particle'' and ``many-particle''
Wigner functions, Wigner norm, and mana.

\subsection{Haar distribution} \label{ss:haar}

The Haar distribution is the (unique) probability distribution on normalized pure states that is invariant under unitaries.
It corresponds to a measure\footnote{One could also use $\delta(1-|\psi|)$, but the difference in power of $|\psi|$ only affects the normalization factor $\mathcal N$.}
\begin{equation}
  d\mu(\psi) = \mathcal N \prod_j \left[\frac {i\; d\psi_j\, d\psi_j^*} 2\right] \delta(1 - |\psi|^2)\;.
\end{equation}
The normalization $\mathcal N$ is
\begin{equation}
  \mathcal N = (d-1)! \pi^{-d}\;;
\end{equation}
we find it more convenient to write this
\begin{equation}
  \mathcal N = \frac{2 \Gamma(d)}{\Gamma(a)\Gamma(b) S_{2a-1} S_{2b-1}}\;,\quad a+b = d
\end{equation}
where $\Gamma$ is the gamma function and $S_{n}$ is the surface area of the unit $n$-sphere (viz., the unit sphere in $\mathbb R^{n+1}$).
We will write
\begin{equation}
  \expcth{f}
\end{equation}
for the average of some function $f(\psi)$ with respect to the Haar distribution.
(For notational convenience we drop the bra-ket notation when discussing arbitrary functions of states.)

In Sec.~\ref{s:mixed}, we consider ``Haar'' distributions on mixed states---that is, distributions invariant under unitaries.
Such a distribution is no longer unique;
to see this, note that the orbits of different diagonal matrices under conjugation by unitaries do not intersect.
Instead, the distribution is specified by its entanglement spectrum.
We can imagine (at least) three such distributions:
\begin{enumerate}
\item \label{it:simp-stat-mix} simple statistical mixtures $\rho = (1 - \alpha) d^{-\ell} I + \alpha \ketbra{\psi}{\psi}$ with $\ket \psi$ from the pure-state Haar distribution,
\item complex statistical mixtures $N^{-1} \sum_j \ketbra{\psi_j}{\psi_j}$ with the $\ket \psi_j$ i.i.d. from the pure-state Haar distribution (roughly infinite-temperature density matrices), and
\item reduced density matrices $\rho_A = \tr_B \ketbra{\psi}{\psi}$ on an $\ell$-site subsystem of an $L$-site system $A \cup B$
\end{enumerate}
where we choose $\alpha, N$, and $L$ respectively to (approximately) fix the entanglement entropy.
The three distributions give density matrices with different spectral properties.
In Sec.~\ref{s:mixed} we alternate between ensembles
1 (simple statistical mixtures)
and 3 (reduced density matrices of pure states on larger systems)
as appropriate:
ensemble 3 is more appropriate for applications and analytical calculations,
but ensemble 1 offers fine-grained control over the entropy.
We also check that all three ensembles give similar results (Sec.~\ref{ss:ensembles}).

\section{Summary of results}\label{s:summary}

For mixed states of second \Renyi entropy $S_2$ on Hilbert spaces of dimension $d^\ell \gg 1$,
\begin{align*}
  \begin{split}
    \expcth{\mathcal W} &= \sqrt{2/\pi}\;(\sigma/\mu) e^{-\mu^2/2\sigma^2} + \erf(\mu/\sigma\sqrt{2})\;.
  \end{split}
\end{align*}
where
\[  \frac {\sigma^2}{\mu^2} = e^{\Delta} - 1\;, \qquad \Delta := \ell \ln d - S_2\;. \]
We call $\Delta$ the \textbf{entropy deficit};
it controls the Wigner norm (and mana) of mixed states
in this asymptotic regime.
The mixed state Wigner norm has two limiting cases:
\begin{align*}
  \expcth{\mathcal W} &= \sqrt{2/\pi}\; e^{\Delta / 2 } + O(e^{-\Delta/2})\;, &\Delta > 1&\; \\
  \expcth{\mathcal W} &= 1 + \sqrt {2/\pi} \Delta^{3/2} e^{- (2\Delta)^{-1} }\left[1 + O(\Delta)\right]\;, &\Delta < 1&\;. \\
\end{align*}
These limits correspond to ``nearly pure'' and ``nearly maximally mixed'' states respectively.

These expressions come from a Gaussian approximation. 
The exact calculation gives a complicated expression \eqref{eq:mixed-characteristic-exact};
it reduces to the Gaussian approximation in the limit $d \gg 1$.

In the pure-state case that asymptotic Gaussian approximation gives
\[ \expcth{\mathcal W}  = \sqrt{2\pi} d^{\ell/2}\qquad\text{(pure state)}\]
while the exact calculation gives
\[ \expcth{\mathcal W} = \frac {d!!}{(d-1)!!} \qquad\text{(pure state)}\;.\]

The variance of the Wigner norm is
\[ \expcth{\Delta W^2} = O(d^{-\ell}). \]
This means that
\begin{equation*}
  \expcth{\mathcal M} \equiv \expcth{\ln \mathcal W} = \ln \expcth{\mathcal W} + O(d^{-\ell})\;,
\end{equation*}
justifying the following expressions for the mana:
\begin{align*}
  \expcth{\mathcal M} &\approx \frac 1 2 [\ell \ln d - \pi/2] &\text{pure state}&\; \\
  \expcth{\mathcal M} &\approx \frac 1 2 [\Delta - \pi/2] &\Delta > 1 &\; \\
  \expcth{\mathcal M} &\approx \sqrt {2/\pi} \Delta^{3/2} e^{- (2\Delta)^{-1} } &\Delta < 1&\;. \\
\end{align*}
All of these expressions are close to
\begin{equation}
  \expcth{\mathcal M} \approx \max\left(0,\ \frac 1 2 [\ell \ln d - S_2 - \ln \pi/2]\right)\;,
\end{equation}
which serves as a useful quick estimate.

\section{Mana of random mixed states}\label{s:mixed}

In this section we compute the mana of random mixed states.
We do so first in an intuitively justified, heuristic Gaussian approximation (Sec.~\ref{ss:mixed-gaussian}).
We then compute the mana in an exact calculation (Sec.~\ref{sec:mixed-exact})
for reduced density matrices of pure Haar states on larger systems (ensemble 3~above).
This exact calculation reduces to the Gaussian calculation in the limit of $d^\ell \gg 1$,
justifying that calculation as a controlled approximation.

Technically, the exact and Gaussian calculations different situations:
in the Gaussian calculation
we fix the system's entropy,
while in the exact calculation
we fix an ancilla dimension
and allow the system's entropy to vary.
We limn the relationship between the two results in Sec.~\ref{s:connect-gaussian-exact}.

The exact calculation treats a specific ensemble,
but the Gaussian calculation is ensemble-agnostic:
its result is specified by the entropy deficit $\Delta$.
In Sec.~\ref{ss:ensembles} we therefore check numerically that for entropy deficit $\Delta \gtrsim 1$ or Hilbert space dimension $d \gg 1$
the three ensembles of \ref{ss:haar} have the same mana.

\subsection{Heuristic Gaussian calculation}\label{ss:mixed-gaussian}

Take a system of $\ell$ qudits, and consider
random mixed states with some fixed second \Renyi entanglement entropy $S_2$.

The average $\expcth{\mathcal W}$ follows from the average of the Wigner function components.
The Wigner function components $W(p,q)$ of a state $\rho$ are unitarily equivalent,
so for Haar-random states they are identically distributed.
Because they are the sum of $d^2 \ge 9$ random variables, namely the matrix elements $A_{ij}\rho_{ij}$,
one expects them to be (roughly) normally distributed
\begin{equation}
  \label{eq:wigner-dist-gaussian}
  p(W) dW \approx \frac 1 {\sigma \sqrt{2\pi} } e^{- W^2 / 2\sigma^2}\;.
\end{equation}
The normalization and entropy conditions Eq.~\eqref{eq:wigner-norm-trace-constraint} give
\begin{subequations}
  \label{eq:wigner-mixed-mu-sigma}
  \begin{align}
    \mu &= d^{-2} \\
    \begin{split}
      \sigma^2 &= d^{-4\ell}\left[e^{\ell \ln d - S_2} - 1\right]\\
      &= d^{-4\ell}\left[e^\Delta - 1\right]
      \end{split}
  \end{align}
\end{subequations}
in terms of the entropy deficit
\begin{equation}
  \label{eq:entropy-deficit}
  \Delta := \ell \log d - S_2\;.
\end{equation}
This is roughly the distribution of $W(p,q)$ across many Haar states.
For $d \sim 1$ the distribution is not quite the Gaussian distribution \eqref{eq:wigner-dist-gaussian}
because the higher cumulants are appreciable.
Additionally, the normalization and purity conditions \eqref{eq:wigner-norm-trace-constraint} are satisfied exactly by the Wigner function for each state,
whereas $d^2$ i.i.d. variables from the normal distribution \eqref{eq:wigner-dist-gaussian} only satisfy them approximately.
Eq.~\eqref{eq:wigner-norm-trace-constraint} therefore imposes a correlation on those Wigner function components.

\subsubsection{Mean}

Taking the $W(p,q)$ i.i.d.\ from the distribution \eqref{eq:wigner-dist-gaussian} and
defining $\mu_{||}$ for notational convenience, we have
\begin{align}
  \begin{split}
    \label{eq:gaussian-mixed-wigner-fn}
  \expcth{|W(p,q)|} \approx \mu_{||} &= \frac 1 {\sigma \sqrt{2\pi} } \int dW\; |W| e^{-(W - \mu) / 2 \sigma^2} \\
  &= \sqrt{2/\pi}\; \sigma e^{-\mu^2/2\sigma^2} \\
  &\qquad + \mu \erf (\mu/\sigma\sqrt{2})\\
  \end{split}
\end{align}
so 
\begin{align}
  \begin{split}
    \label{eq:gaussian-mixed}
    \expcth{\mathcal W} &= d^{2\ell} \expcth{|W(p,q)|}\\
    &= \sqrt{2/\pi}\;(\sigma/\mu) e^{-\mu^2/2\sigma^2} + \erf(\mu/\sigma\sqrt{2})\;.
  \end{split}
\end{align}
This depends only on the ratio $\sigma / \mu$, but by Eq.~\eqref{eq:wigner-mixed-mu-sigma}
\begin{equation}
  \label{eq:sigma-mu-Delta}
  \frac {\sigma^2}{\mu^2} = e^{\Delta} - 1\;.
\end{equation}
In this approximation, the Wigner norm and mana depend only on this entropy deficit $\Delta$.

\begin{figure}[t]
  \begin{minipage}{0.45\textwidth}
    \includegraphics[width=\textwidth]{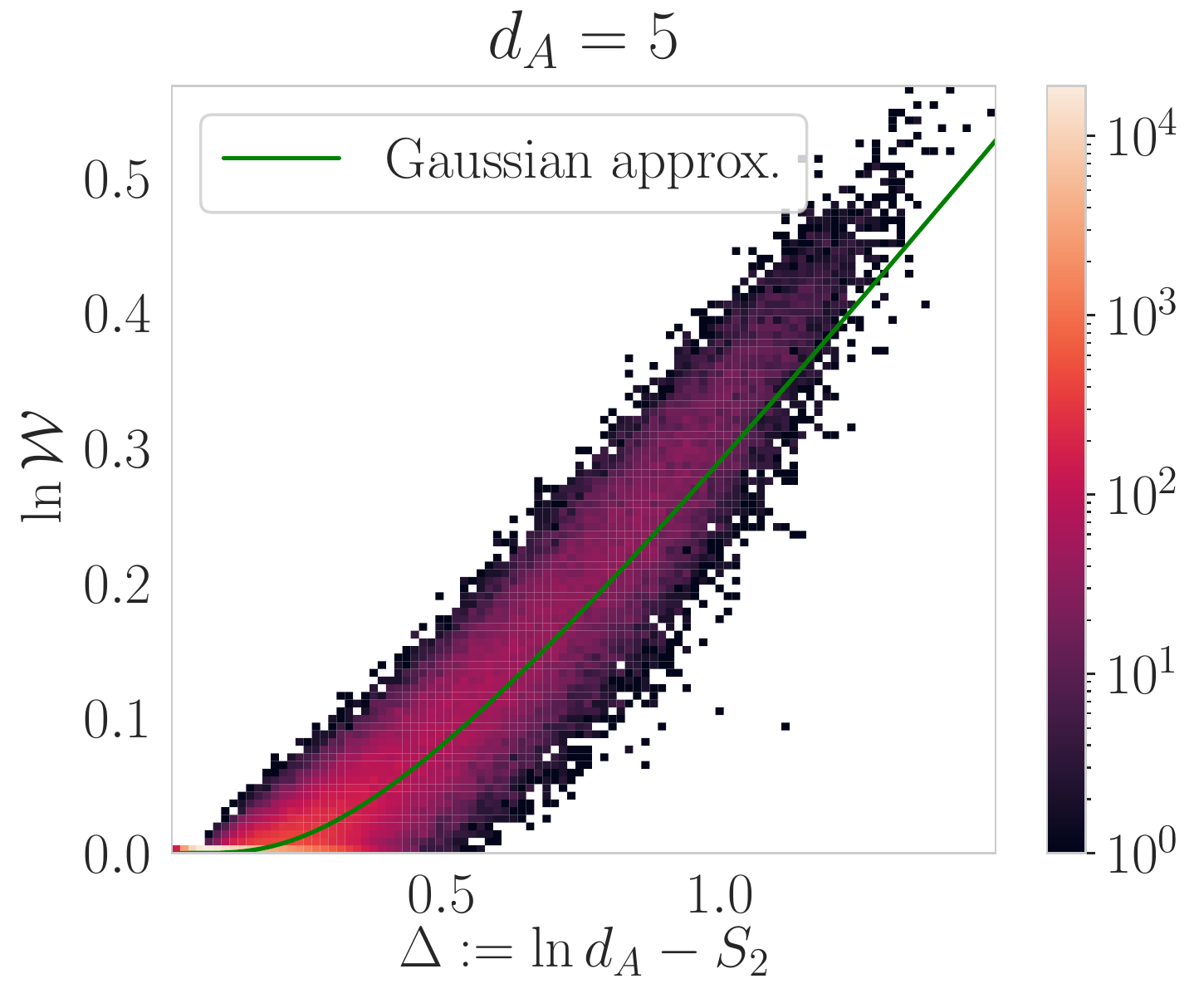}
  \end{minipage}
  
  \begin{minipage}{0.45\textwidth}
    \includegraphics[width=\textwidth]{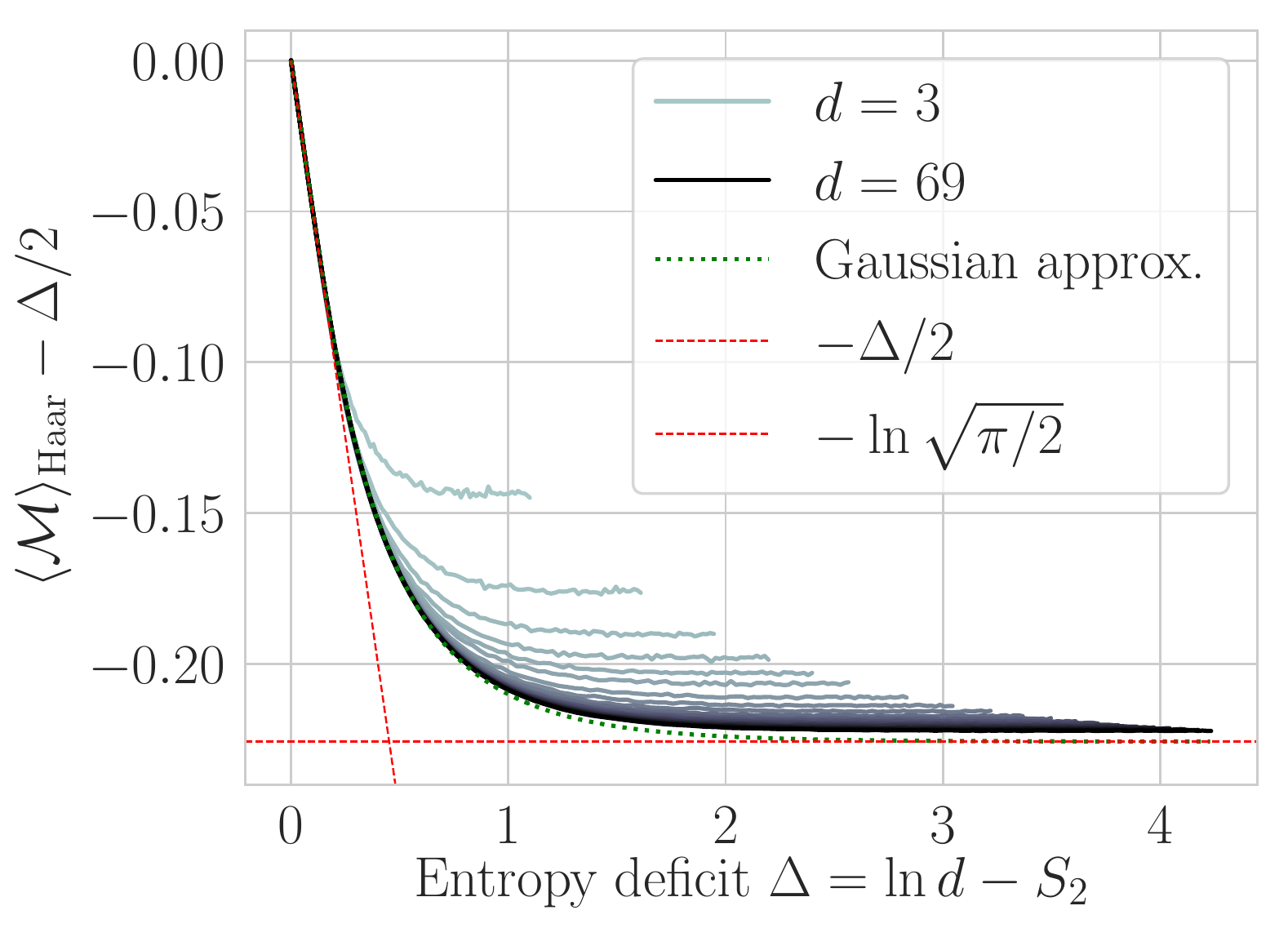}
  \end{minipage}

  \caption{
    \textbf{Random mixed state mana} as a function of entropy deficit,
    together with the Gaussian prediction of \eqref{eq:gaussian-mixed} for the mean.
    \textbf{Top:} Histogram of mana for reduced density matrices
    of random states with system dimension $d_A = 5$
    and ancilla dimension $d_B$ prime in $3 \dots 71$,
    with Gaussian prediction (green line).
    We take 100 random states at each ancilla dimension.
    \textbf{Bottom:} Numerical mean mana for simple statistical mixtures for $d_A \in \{3,5,7,9,11,13,17,21,\dots,69\}$.
    This is the data of Fig.~\ref{fig:mixed-intro},
    replotted to show finite-size corrections and asymptotes more clearly.
    We take 4000 random states at each entropy;
    the variability of the curve suggests the (small) error in our average.
    In both plots, we see that the Gaussian approximation misses a small positive finite-size correction.
  }
  \label{fig:mixed-detail}
\end{figure}

The expressions \eqref{eq:gaussian-mixed} and \eqref{eq:sigma-mu-Delta}
together give the average mana.
We can simplify them for $\Delta > 1$ and $\Delta < 1.$
For $\Delta > 1$, we have $\mu/\sigma \ll 1$,
and a Taylor series expansion of \eqref{eq:gaussian-mixed} gives (at leading order)
\begin{equation}
  \label{eq:wigner-delta->-1}
  \expcth{\mathcal W} = \sqrt{2/\pi} \sigma/\mu = \sqrt{2/\pi}\; e^{\Delta/ 2 }\;.
\end{equation}
Since the variance of the Wigner norm is small (Sec.~\ref{sss:mixed-gauss-variance}),
\begin{equation}
  \expcth{\mathcal M} \approx \ln \expcth{\mathcal W}  = \frac 1 2 [\Delta - \ln \pi/2]\;.
\end{equation}
For $\Delta < 1$, we have $\mu/\sigma \gg 1$,
and the asymptotic expansion of erfc gives
\begin{align}
  \expcth{\mathcal W} &\approx  1 + \sqrt{2/\pi}\; \frac {\sigma^3}{\mu^3} e^{-\mu^2/2\sigma^2} \notag\\
                      &\approx 1 + \sqrt {2/\pi} \Delta^{3/2} e^{- (2\Delta)^{-1} }\;.
\end{align}
and
\begin{align}
  \mathcal M \approx \sqrt{2/\pi} \Delta^{3/2} e^{-(2\Delta)^{-1} }\;.
\end{align}

We plot the mana together with the Gaussian prediction of \eqref{eq:gaussian-mixed} in Fig.~\ref{fig:mixed-detail}.
We see that the Gaussian approximation largely captures the true behavior of the average mana,
but it misses a finite-subsystem correction.

\subsubsection{Variance} \label{sss:mixed-gauss-variance}

\begin{figure}
  \begin{minipage}{0.45\textwidth}
    \includegraphics[width=\textwidth]{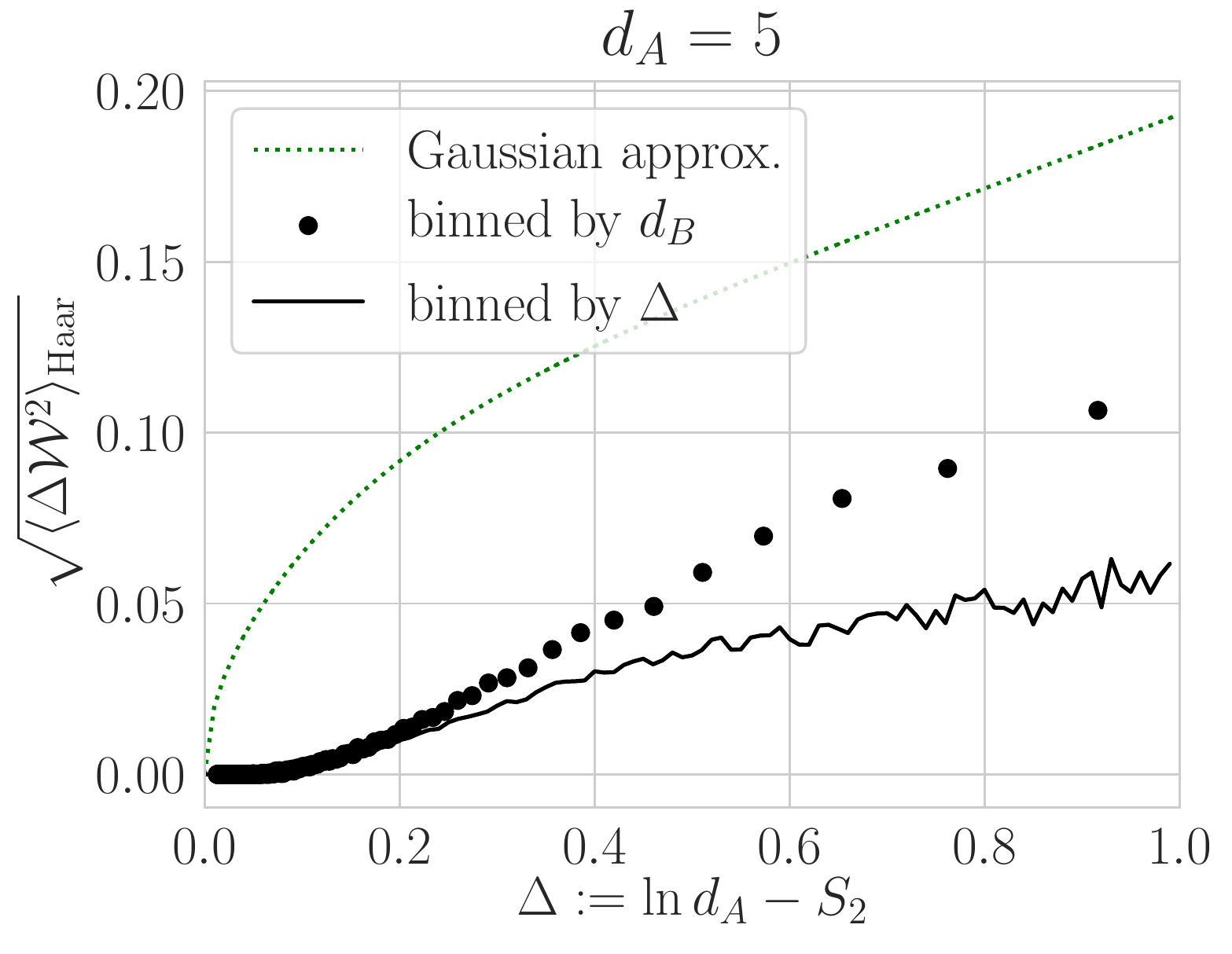}
  \end{minipage}
  
  \caption{ Standard deviation of the Wigner norm
    as a function of entropy deficit $\Delta$
    for Hilbert space dimension $d = 5$.
    The line shows the standard deviation in the Wigner norm
    when states are binned by their entanglement entropy
    (or equivalently entropy deficit);
    points show the standard deviation when states are binned by ancilla dimension $d_B$.
    (The points are assigned the entropy deficit corresponding to \eqref{eq:tr-rho2}.)
    The green dotted line shows the Gaussian approximation \eqref{eq:gauss-mixed-var}
    Compare Fig.~\ref{fig:pure-wigner-norm-std},
    which shows the pure-state special case.
    }
  \label{fig:mixed-var}
  
\end{figure}

The Gaussian approximation predicts a variance
\begin{align}
  \begin{split}
    &\expcth{|W(p,q)|^2} - \expcth{|W(p,q)|}^2 = \sigma^2 - \mu_\|^2
  \end{split}
\end{align}
for the individual $W(p,q)$; the central limit theorem then gives
\begin{align}
  \expcth{\Delta \mathcal W^2} = d^{-2\ell}(\sigma^2 - \mu_\|^2)\;.
\end{align}
Using the entropy result \eqref{eq:wigner-mixed-mu-sigma} for $\sigma$
and the fact that $\expcth{\mathcal W} = d^{2\ell} \mu_\|$,
\begin{align}
  \label{eq:gauss-mixed-var}
  \expcth{\Delta \mathcal W^2} = d^{-2\ell}[e^{\Delta} - 1 - \expcth{\mathcal W}^2]\;.
\end{align}
For $\Delta > 1$,
\begin{equation}
  \expcth{\Delta \mathcal W^2} \approx d^{-2\ell}e^{\Delta} [1 - 2/\pi]\;,
\end{equation}
while, for $\Delta < 1$,
\begin{equation}
  \expcth{\Delta \mathcal W^2} \approx d^{-2\ell}\Delta\;.
\end{equation}

We plot the Gaussian prediction \eqref{eq:gauss-mixed-var}
together with numerical estimates of the standard deviation
in Fig.~\ref{fig:mixed-var}.
For $\Delta \gtrsim 0.5$ we see that the Gaussian prediction has qualitatively correct scaling in both $\Delta$ (top plot) and Hilbert space dimension $d$ (bottom plot),
but overestimates the standard deviation by a constant factor.
This is because choosing the Wigner function components i.i.d. from the Gaussian distribution of Eq.~\eqref{eq:wigner-dist-gaussian}
relaxes the exact constraint
\[ \sum W(p,q)^2 = e^{-S_2} \]
to an approximate result
\[ \sum W(p,q)^2 = e^{-S_2} + \text{[random error]}\: \]
essentially, the Gaussian approximation allows the entropy to vary.
In some situations this is appropriate (cf.\ Sec.~\ref{s:connect-gaussian-exact},
but it results in an overestimate if we bin states by entropy and ask about the variability within each bin,
as we do in Fig.~\ref{fig:mixed-var}.

For $\Delta < 1$, we must additionally contend with the fact that the Gaussian approximation relaxes the exact constraint
\[ \sum W(p,q) = 1 \]
to an approximate result
\[ \sum W(p,q) = e^{-S_2} + \text{[random error]}\;. \]
(This is not important for $\Delta > 1$,
because $\sigma$ controls Eq.~\eqref{eq:gaussian-mixed-wigner-fn}.
$\mu$ appears as either a correction or a convenient way of writing a dimensionality factor.)

\subsection{Exact calculation}\label{sec:mixed-exact}

In this section,
we exactly compute the average mana for reduced densities of Haar pure states.
This is ensemble 3 in the typology of Sec.~\ref{ss:haar}.

Call the system of interest $A$ and its Hilbert space dimension $d_A$;
add an ancilla $B$ with dimension $d_B$.
We wish to calculate the mana of $A$.

More generally we seek the characteristic function
\begin{equation}
  \expcth{e^{z|W(p,q)|}} 
\end{equation}
where we understand the Haar average to be an average over reduced density matrices on $A$
of Haar-distributed states on $A\otimes B$.
Because the $W(p,q)$ are unitarily equivalent,
all the different $W(p,q)$ have identical characteristic functions.
With this characteristic function in hand
we are able to read off
\begin{equation}
  \expcth{|W(p,q)|}\;;
\end{equation}
the Wigner norm is then
\begin{equation}
  \expcth{\mathcal W} = \sum_{pq} \expcth{|W(p,q)|} = d_A^2 \expcth{|W(p,q)|}\;.
\end{equation}

The calculation is somewhat involved;
here we give first the results and then the calculation.

\subsubsection{Results}

\begin{figure}[t]
  \includegraphics[width=0.45\textwidth]{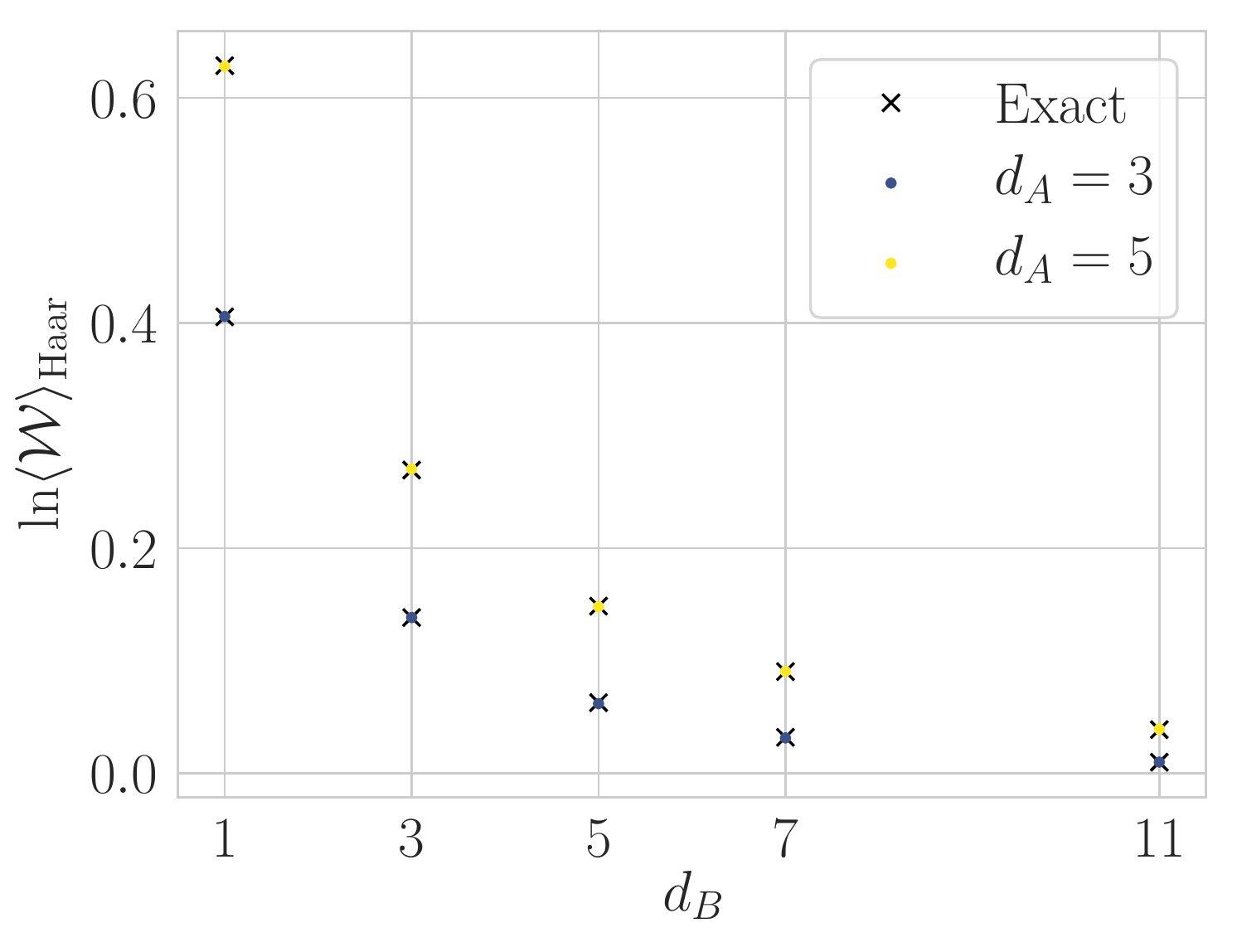}
  \caption{\textbf{Mana of reduced density matrices} as a function of ancilla dimension $d_B$.
    Colored dots show numerical results for system dimension $d_A = 3,5$;
    exes show the exact calculation \eqref{eq:mixed-wigner-exact}.}
  \label{fig:exact-mana-dB}
\end{figure}

We find that the characteristic function is
\begin{widetext}
  \begin{multline}
    \label{eq:mixed-characteristic-exact}
    \characteristic = \frac{\Gamma(D)}{2^{D-1}\Gamma(a)\Gamma(b)}\sum_{n=0}^{\infty} \frac{n!}{d^{n}}\Big[ \frac{\Gamma(b)}{\Gamma(1+b+n)} {}_{2}F_{1} (1-a,1+n;1+b+n;-1) \\ +\frac{\Gamma(a)}{\Gamma(1+a+n)} {}_{2}F_{1} (1-b,1+n;1+a+n;-1) \Big] \frac{z^{n}}{n!}
  \end{multline}
where $D=d_{A}d_{B}$, $a=d_{B}\frac{d_{A}+1}2$, and $b=d_{B}\frac{d_{A}-1}2$.
The first-order term in this series gives $\expcth{|W(p,q)|}$,
hence the Wigner function;
the result is conveniently written in terms of multinomial coefficients
\begin{equation}
  \label{eq:mixed-wigner-exact}
  \expcth{\mathcal W} = \frac {d_A}{2^{D-1}} \binom{D}{(a-1),(b-1),1}
  \left[
    \frac 1 {b (b+1)} {}_2F_{1}(1-a, 2, 2+b, -1)  
    +
    \frac 1 {a (a+1)} {}_2F_{1}(1-b, 2, 2+a, -1)
  \right]\;.
\end{equation}
\end{widetext}
We plot this prediction together with numerical samples in Fig.~\ref{fig:exact-mana-dB}.

\subsubsection{Calculation}

Let
\begin{equation}
  \ket{\psi} = \sum_{i=1}^{d_{A}}\sum_{j=1}^{d_{B}}  \psi_{ij} \ket{i}\otimes \ket{j}\;.
\end{equation}
be a Haar-random state on $A\otimes B$,
and take the reduced density matrix
\begin{equation}
  \rho_A = \tr_B \ketbra{\psi}{\psi}= \sum_{i,i'=1}^{d_{A}} \sum_{j=1}^{d_{B}} \psi_{ij}\psi_{i'j}^{*} \ketbra{i}{i'}.\;.
\end{equation}
Since the $A(p,q)$ are unitarily equivalent,
we can pick a $p,q$ and work in the eigenbasis of that $A(p,q)$;
the Wigner function is then
\begin{equation}
  \label{eq:wigner-function-ancilla}
  W(p,q) = d_{A}^{-1}\left(\sum_{i=1}^{\frac{d_{A}+1}2}\sum_{j=1}^{d_{B}}|\psi_{ij}|^{2} - \sum_{i=\frac{d_{A}+3}2}^{d_{A}}\sum_{j=1}^{d_{B}}|\psi_{ij}|^{2}  \right)\;.
\end{equation}
This is the quantity whose characteristic function $\expcth{e^{z|W(p,q)|}}$ we seek.

To proceed, we break up the Hilbert space $A \otimes B$
by the sign of each dimension in Eq.~\eqref{eq:wigner-function-ancilla}:
that is, we split a wavefunction $\ket \psi$ on $A \otimes B$
into $\ket \psi = \ket \phi \oplus \ket \xi$
such that
\begin{equation}
  W= d_{A}^{-1}( |\phi|^{2}-|\xi|^{2} )
\end{equation}
with $\phi$ in $d_{B}\frac{d_{A}+1}2$ complex dimensions
and $\xi$ in $d_{B}\frac{d_{A}-1}2$ complex dimensions.
Defining
\begin{align*}
  a &:= d_B \frac{d_{A} + 1} 2\\
  b &:= d_B \frac{d_{A} - 1} 2
\end{align*}
for the dimensions of the positive and negative spaces respectively,
the Haar measure breaks up into
\begin{equation}
 d\mu(\psi) = \mathcal N \prod_{j=1}^{a}\frac{i d\phi_{j} d\phi_{j}^{*}}{2} \prod_{k=1}^b  \frac{i d\xi_{k} d\xi_{k}^{*}}{2}\delta(1-|\phi|^{2}-|\xi|^{2})
\end{equation}
with normalization
\begin{equation}
  \label{eq:haar-norm-ab}
  \mathcal N = \frac{2 \Gamma(d)}{\Gamma(a)\Gamma(b) S_{2a-1} S_{2b-1}}
\end{equation}
and the characteristic function is
\begin{equation}
  \expcth{e^{z|W(p,q)|}} = \int d\mu(\phi\oplus\xi) \, e^{z d_{A}^{-1}| |\phi|^{2}-|\xi|^{2}|}\;.
\end{equation}
The angular integrals about $\phi$ and $\xi$ cancel the $S_{2a-1} S_{2b-1}$ in the denominator of Eq.~\eqref{eq:haar-norm-ab} and---defining compact notation $G(z)\equiv \expcth{e^{z|W(p,q)|}}$ for the characteristic function---
\begin{align}
  \begin{split}
    G(z) = \frac{2\Gamma(D)}{\Gamma(a)\Gamma (b)} \int_{0}^{1}&ds\int_{0}^{1}dr\\
    &\times\delta(1-r^{2}-s^{2}) e^{z d_{A}^{-1}|r^{2}-s^{2}|}\;.
    \end{split}
\end{align}
 % In order to now take the average over Haar states, we write out
% \begin{equation}
%   \langle e^{z|W(p,q)|}\rangle = \frac{2 \Gamma(D)}{\Gamma(d_{B}\tfrac{d_{A}+1}2) \Gamma(d_{B}\tfrac{d_{A}-1}2)} \int_{0}^{1} r^{d_{B}(d_{A}+1)-1}(1-r^{2})^{d_{B}\frac{d_{A}-1}2 - 1} e^{z d^{-1} | 2r^{2}-1|} dr
% \end{equation}
Integrating over $s$ gives
\begin{equation}
  G(z) = \frac{2 \Gamma(D)}{\Gamma(a) \Gamma(b)} \int_{0}^{1} r^{2a-1}(1-r^{2})^{b - 1} e^{z d_A^{-1} | 2r^{2}-1|} dr;
\end{equation}
after a substitution $x=2r^{2}-1$ this becomes.
\begin{equation}
  \label{eq:exact-integral}
  G(z) = \frac{ \Gamma(D)}{\Gamma(a) \Gamma(b)2^{D-1}} \int_{-1}^{1} (1+x)^{a-1}(1-x)^{b-1} e^{z d_A^{-1}|x|} dx.
\end{equation}
We evaluate the integral by breaking the range into two parts $-1 \le x \le 0$ and $0 \le x \le 1$ related by the swap $a \leftrightarrow b$,
Taylor expanding the exponential,
and appealing to \cite{gradshtein_table_2007};
this gives Eq.~\eqref{eq:mixed-characteristic-exact}.

\subsection{Connecting the Gaussian approximation and the exact calculation} \label{s:connect-gaussian-exact}

Instead of evaluating the integral \eqref{eq:exact-integral} exactly,
we can estimate it by a saddle point approximation.
The result is a Gaussian integral.
Writing $W := d^{-1} x$ for the Wigner function,
\begin{align}
  \label{eq:saddle-point-char}
  \begin{split}
    &\characteristic\\
    &\qquad =\frac 1 {\sigma \sqrt{2\pi} } \int_{-\infty}^\infty dW\; e^{-\frac{(W-\mu)^2}{2\sigma^2}} e^{z |W|} \\
    &\qquad\qquad + O[(d_A d_B)^{-1}]
  \end{split}
\end{align}
with
\begin{subequations}
    \label{eq:saddle-params}
  \begin{align}
    \mu &= \frac{d_B}{d_A(D-2)} = \frac 1 {d_A^2} + O[(d_Ad_B)^{-1}]\\
    \begin{split}
      \sigma^2 &= \frac{1}{d_A^2(D-2)}\\
      &= d_A^{-4}\left[d_A\expcth{\tr \rho_A^2} - 1\right] +  O[(d_Ad_B)^{-1}]\;,
    \end{split}
  \end{align}
\end{subequations}
where we use that
\begin{equation}
  \expcth{\tr \rho_A^2} = \frac {d_A + d_B} {d_A d_B + 1}\;.
\end{equation}
(The Gaussian approximation for the integrand and the Sterling's formula approximation for the normalization factors each induce an error.
Extending the limits of integration induces an exponentially small error.)
Note that the error is controlled by $d_A d_B$:
this approximation holds for either $d_A \gg 1$ or $d_B \gg 1$, but does not require both.

The calculation leading to \eqref{eq:saddle-point-char} does not use the absolute value $|W|$.
So $W$ has a similar characteristic function \eqref{eq:saddle-point-char}, dropping the absolute value;
since the characteristic function determines the probability distribution we have shown that---in the limit of large dimension $d_A d_B$---the Wigner function is distributed as a Gaussian with mean and variance \eqref{eq:saddle-params}.
Furthermore, computing corrections follows the standard procedure for saddle point approximation schemes.

The mean and variance \eqref{eq:saddle-params}
give that the Gaussian distribution satisfies
the normalization and entropy constraints \eqref{eq:wigner-norm-trace-constraint} at leading order.
One might worry that the $O[(d_Ad_B)^{-1}]$ corrections to $\mu, \sigma^2$
mean that the distribution does not satisfy those constraints exactly.
But the distribution is only Gaussian at leading order; subleading corrections to the distribution will cancel the subleading terms in \eqref{eq:saddle-params}.

In some sense, then, the heuristic Gaussian calculation of Sec.~\ref{ss:mixed-gaussian} consists of taking
the saddle point approximation \eqref{eq:saddle-point-char}
with \eqref{eq:saddle-params} seriously,
even for $d_A, d_B \gtrsim 1$.
But one must be careful.
The exact calculation does not straightforwardly provide finite-dimension corrections to the Gaussian calculation
because the Gaussian and exact calculations treat different situations.
The Gaussian calculation considered normalized states with given second Renyi entropy $S_2$;
the entropy controlled the variance of the Gaussian, hence the mana. 
But in the exact calculation we do not fix the entropy.
Rather, we entangle the system of interest with an ancilla of fixed dimension,
and choose a random state on the tensor product space.
The resulting average purity is 
\begin{equation}
  \label{eq:tr-rho2}
  \expcth{\tr \rho_A^2} = \frac {d_A + d_B} {d_A d_B + 1}\;,
\end{equation}
and for $d_A \gg 1$ large almost all states will have purity very close to this average.
But for $d_A \sim 1$ there will be appreciable variation,
and reduced density matrices will be more broadly distributed in the generalized Bloch ball
than random states of fixed entropy,
and the variance of the Wigner norm will consequently be greater.

\begin{figure}
  \includegraphics[width=0.45\textwidth]{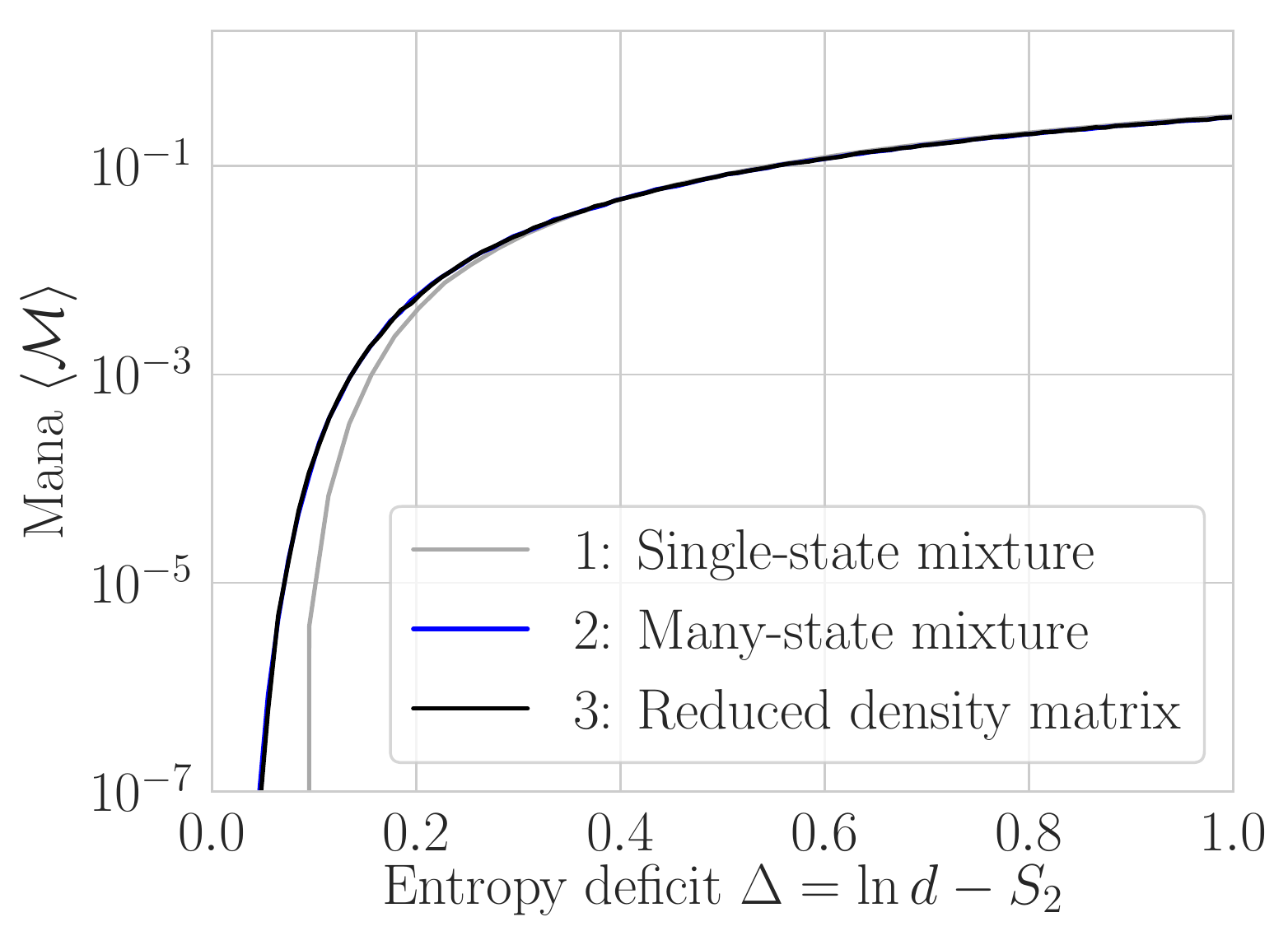}
  \caption{
    \textbf{Open-system mana in three distributions:}
    a single-state statistical mixtures $\rho = d^{-1}(1 - \alpha) I + \alpha \ketbra{\psi}{\psi}$,
    a many-state statistical mixture $\rho = N^{-1}\sum_{j = 1}^N \ketbra{\psi_j}{\psi_j}$,
    and a $d$-dimensional subsystem of a larger system $\rho = \tr_{B} \ketbra{\psi}{\psi}$.
    The many-state statistical mixture (2)
    and the reduced density matrix (3)
    turn are identical, because their spectral properties are very similar.
    In each case we average over the states near each entropy.
    The (single-site) Hilbert space dimension is $d = 11$.
    }
  \label{fig:mana-ensemble-comparison}
\end{figure}

Fig.~\ref{fig:mixed-var} shows how
the variance of the Wigner norm differs
when one considers constant ancilla dimension
as opposed to constant entropy.
We can see that the variance is greater when states are binned by $d_B$;
this confirms that entropy---not ancilla dimension $d_B$---is the appropriate knob.

\subsection{Comparing ensembles of random mixed states}\label{ss:ensembles}

Fixing the system size and entropy does not specify the distribution (cf.\ \ref{ss:haar}),
but it does specify the two inputs---normalization and entropy---to the heuristic Gaussian calculation below.
This suggests that all distributions invariant under unitaries will give the same average mana,
at least for large Hilbert space dimension.
Numerical evidence (Fig.~\ref{fig:mana-ensemble-comparison})
confirms that this is the case.
Many-state statistical mixtures and the reduced density matrices behave identically for all $\Delta$,
and the simple statistical mixture is very similar
for $\Delta \gtrsim 1/2$.
For small entropy deficit $\Delta$
(that is, states close to the maximally-mixed state),
the simple statistical mixture begins to differ from the other two distributions.
This is because the region of zero Wigner norm
has a complicated geometry---especially near the identity matrix,
where a state may be near many facets of the convex hull of the stabilizer states.
The three distributions probe this geometry in slightly different ways.
As $d$ increases the region in which the complicated geometry is important
occupies a smaller and smaller ball around the maximally mixed state at $\Delta = 0$,
so one would expect the regime in which the three ensembles agree
to extend closer and closer to $\Delta = 0$.
We have checked that this is the case.

\section{Mana in random pure states}\label{s:pure}
In Sec.~\ref{s:mixed} we computed the mana for random mixed states of arbitrary entropy,
but many applications (e.g.\ the study of approximate unitary designs) will require pure states.
We therefore investigate the pure state special case in somewhat more detail in this section.

\subsection{Heuristic Gaussian calculation}\label{ss:pure-gaussian}

\begin{figure}
  \includegraphics[width=0.45\textwidth]{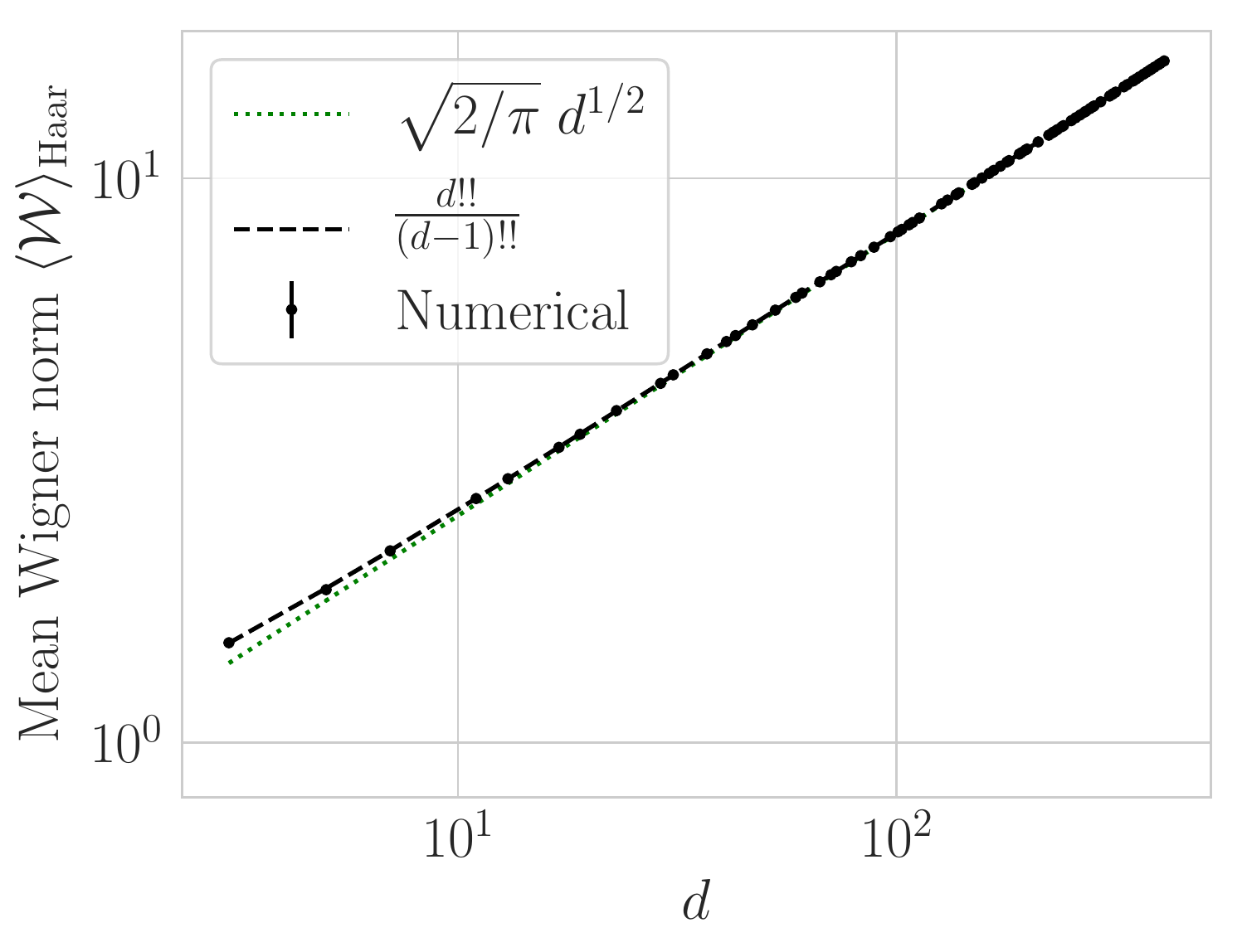}
  \caption{
    \textbf{Haar-averaged pure state Wigner norm}
    showing numerical estimates (black dots),
    the exact calculation Eq.~\eqref{eq:pure-exact-1} of Sec.~\ref{ss:pure-exact} (black dashed line),
    and the Gaussian approximation Eq.~\eqref{eq:gaussian-mixed-wigner-fn} of Sec.~\ref{ss:pure-gaussian} (green dotted line).
    Statistical error in the numerical estimates is smaller than the dots.
  }
  \label{fig:pure-wigner-norm}
\end{figure}

\begin{figure}
  \includegraphics[width=0.45\textwidth]{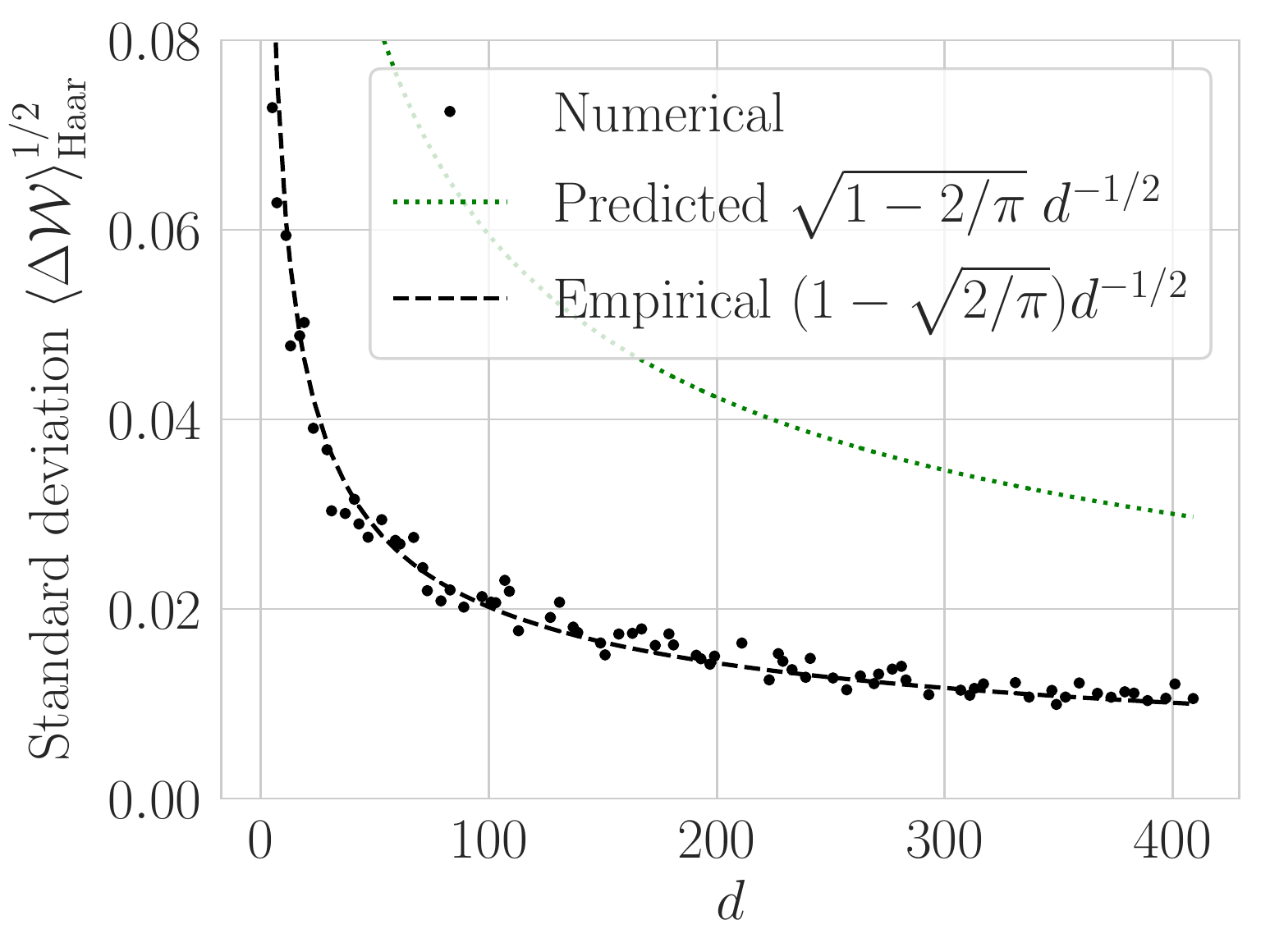}
  \caption{
    \textbf{Standard deviation of Wigner norm}
    over Haar distribution showing numerical estimates (black dots),
    the prediction Eq.~\eqref{eq:g-var} of the Gaussian approximation (green dotted line),
    and an empirical estimate (black dashed line).
    }
  \label{fig:pure-wigner-norm-std}
\end{figure}

\begin{figure}
  \begin{minipage}{0.45\textwidth}
    \includegraphics[width=\textwidth]{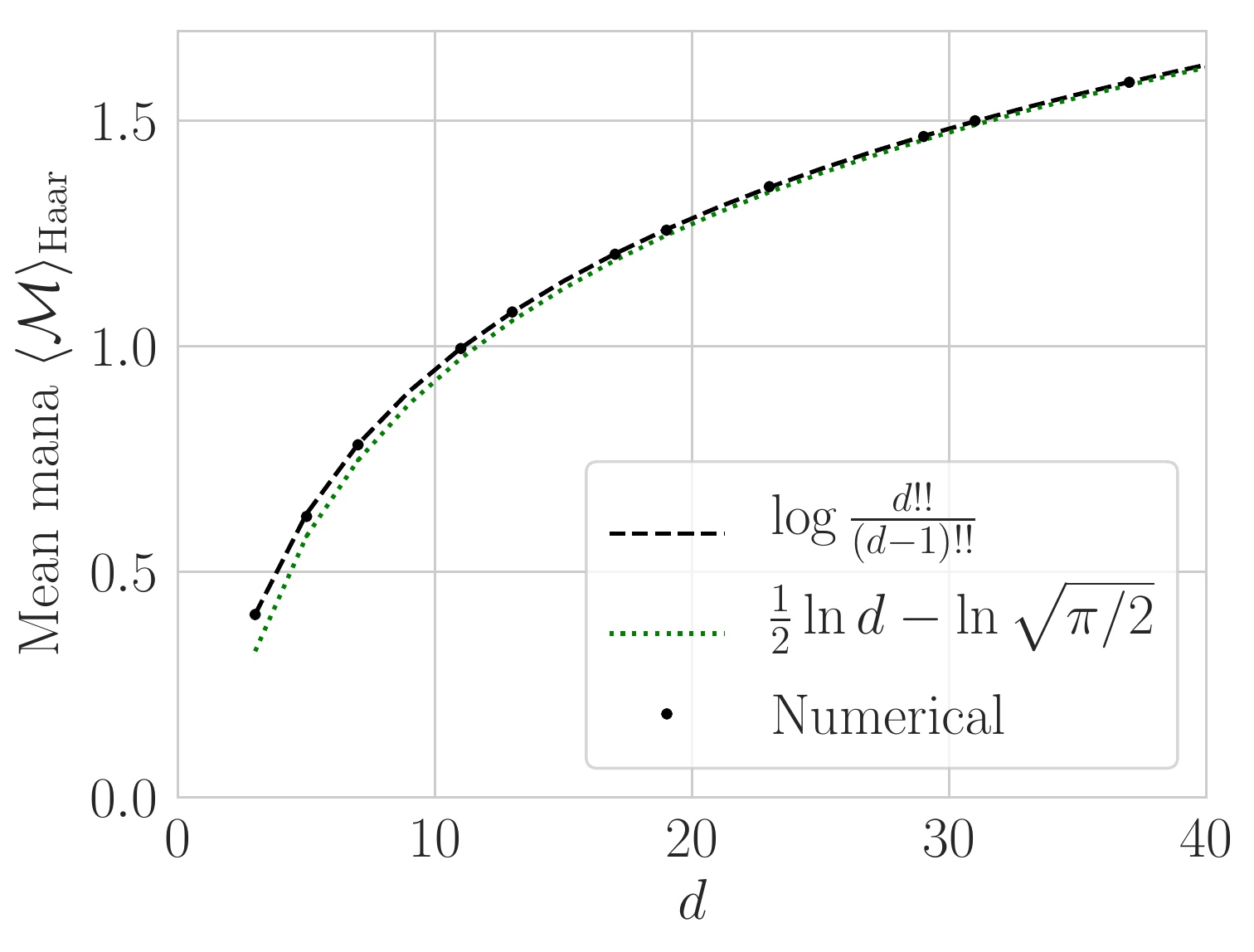}
  \end{minipage}

  \begin{minipage}{0.45\textwidth}
    \includegraphics[width=\textwidth]{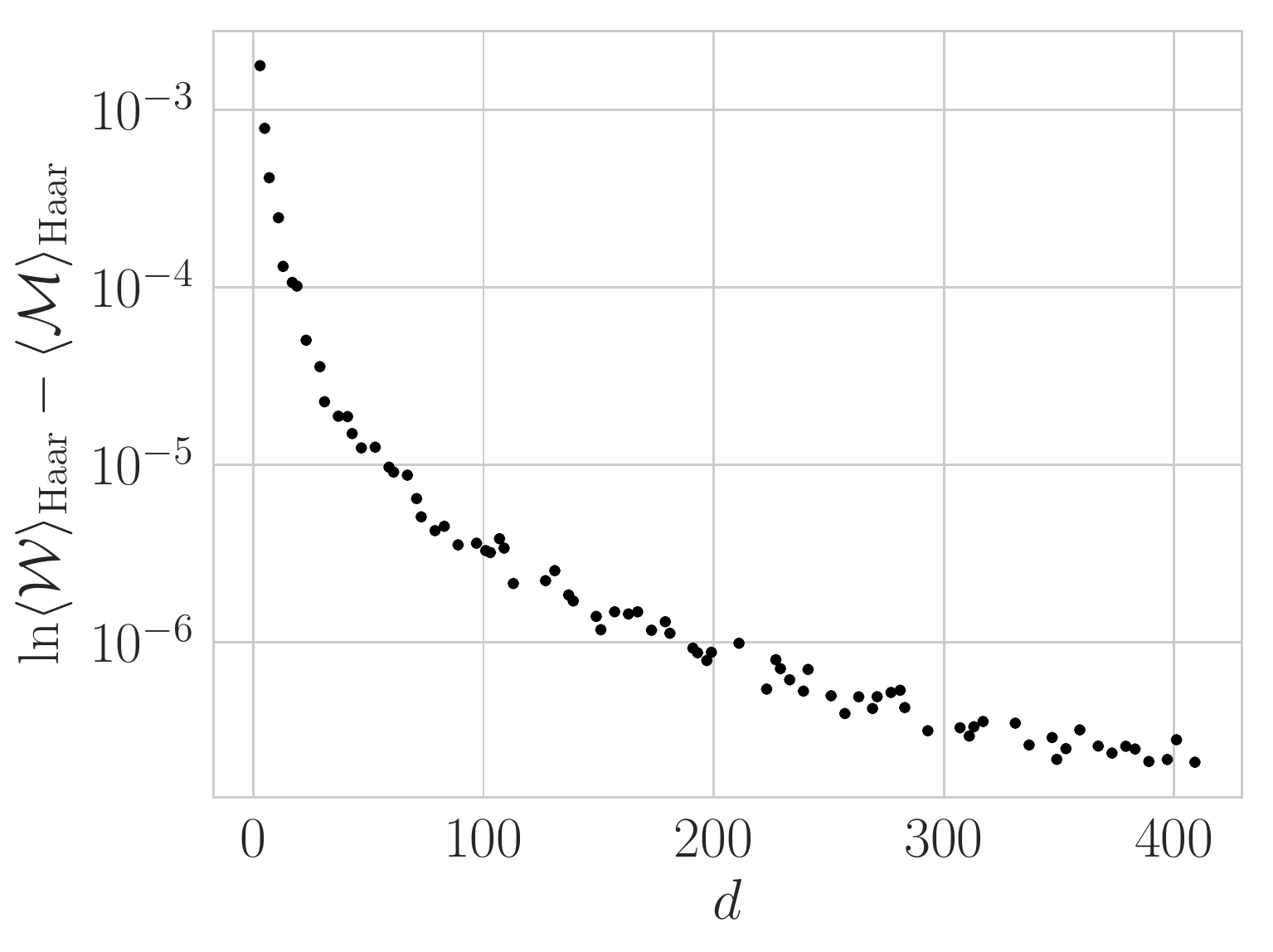}
  \end{minipage}
  
  \caption{\textbf{Haar-averaged pure state mana}.
    \textbf{Top}:
    the numerical estimates (black dots) and the log of the two predicted mean Wigner norms.
    \textbf{Bottom}: deviation of the mean mana from the log of the mean Wigner norm.
    Because the standard deviation is small (Fig.~\ref{fig:pure-wigner-norm-std}),
    this deviation is very small.
  }
  \label{fig:pure-mana}
\end{figure}

Pure states have entropy deficit
\begin{equation}
  \Delta = \ell \ln d > 1\;.
\end{equation}
The Gaussian approximation \eqref{eq:wigner-delta->-1} therefore gives
\begin{equation}
  \expcth{\mathcal W} = \sqrt{2/\pi}\; d^{\ell/2}\;;
\end{equation}
we plot this in Fig.~\ref{fig:pure-wigner-norm},
together with the exact result of Sec.~\ref{ss:pure-exact}
and the average of a numerical sample.

The assumptions leading to these estimates of $\expcth{\mathcal W}$
again give an estimate of the variance $\expcth{\Delta\mathcal W^2}$.
We have assumed that the Wigner function components $W(p,q)$ have mean and variance $\mu, \sigma^2$ given by the normalization and entropy,
in this case
\begin{subequations}
  \begin{align}
    \mu &= d^{-2} \\
    \sigma^2 &= d^{-4\ell}\left[d^{\ell} - 1\right]\;,
  \end{align}
\end{subequations}
so the variance of their absolute value is
\begin{equation}
  \sigma^2 - \mu_{||}^2 = d^{-3}(1 - 2/\pi)\;.
\end{equation}
We have further taken the Wigner function components \textit{of a single state} to be i.i.d. with this mean and variance.
In this approximation, the central limit theorem gives
\begin{equation}
  \label{eq:g-var}
  \expcth{\Delta \mathcal W^2} \approx d^{-1}(1 - 2/\pi)\;.
\end{equation}
We can see from the numerics (Fig.~\ref{fig:pure-wigner-norm}) that \eqref{eq:g-var} has the right scaling,
but is off by a constant factor.
As in Sec.~\ref{sss:mixed-gauss-variance},
the discrepancy is due to the correlations between the $W(p,q)$ that we ignored when we took them i.i.d.
Because $\sum_{p,q} W(p,q) = 1$ and $\sum_{p,q} W(p,q)^2 = d^{-1}$ exactly,
the variation in $ \mathcal W = \sum_{p,q} |W(p,q)|$ between states is smaller.
Eq.~\eqref{eq:g-var} is therefore an overestimate of the variation in Wigner norm.

Because most states have Wigner norm very close to the average $\expcth{\mathcal W}$,
\begin{align}
  \expcth{\mathcal M} &\equiv \expcth{\ln \mathcal W} \approx \ln \expcth{\mathcal W}\\
  &\approx \frac 1 2 \ln d - \ln \sqrt{\pi/2}\;:
\end{align}
that is,
the mana of a Haar random state has the maximal mana \eqref{eq:jensens-bound}, up to a constant correction $\ln \sqrt{\pi/2}$.
In Fig.~\ref{fig:pure-mana} we plot the average mana $\expcth{\mathcal M}$ from numerics,
together with the predicted $\ln \expcth{\mathcal W}$ of the heuristic Gaussian calculation
and the exact calcuation of Sec.~\ref{ss:pure-exact}.
We also check that
\begin{equation}
 \ln \expcth{\mathcal W} - \expcth{\mathcal M}
\end{equation}
is small by plotting its average for numerical samples from the Haar distribution (Fig.~\ref{fig:pure-mana} bottom).

\subsection{Exact $(2d)$-sphere calculation}\label{ss:pure-exact}
For pure states the characteristic function \eqref{eq:mixed-characteristic-exact} becomes
\begin{subequations}
\begin{align}
  &\expcth{e^{z |W(p,q)|} } \notag \\
  \begin{split}
    &\quad=
    \frac{\Gammalr{\frac d 2}     \Gammalr{\frac{d+1} 2} }
         {\Gammalr{\frac {d+1} 4} \Gammalr{\frac{d+3} 4} }
         \\&\qquad
         \times\sum_{n = 0}^\infty
         \frac{
           \Gammalr{\frac{d+1} 4 + \frac n 2}
           \Gammalr{\frac{d+3} 4 + \frac n 2}
         }
         {
           \Gammalr{\frac n 2 + 1}
           \Gammalr{\frac{d + n} 2}
         }
         \frac{z^n}{d^n}
       \end{split}
  \label{eq:pure-exact-series}
  \\
  \begin{split}
    &\quad= B\left(\frac{d+1} 4, \frac{d+1} 3\right)
    \\ &\qquad
    \times \Xi_1\left(\frac{d-1} 2, 1, \frac{3-d} 2, \frac{d+1} 2; \frac 1 2, \frac z d\right)\;.
  \end{split}
         \label{eq:pure-exact-hypergeo}
\end{align}
\end{subequations}
where $B$ is the Beta function and $\Xi_1$ is a Humbert hypergeometric function.
From the series expression \eqref{eq:pure-exact-series} we can read off the first two moments
\begin{subequations}
  \begin{align}
    \expcth{|W(p,q)|} &= \frac 1 {d^2} \frac{ d!! } {(d-1)!!} \label{eq:pure-first-moment}\\
    \expcth{|W(p,q)|^2} &= d^{-3}\;.
  \end{align}
\end{subequations}
For $d \gg 1$ the first moment becomes
\begin{equation}
  \label{eq:pure-exact-1}
  \expcth{|W(p,q)|} \approx d^{-2} \sqrt{2/\pi}\; d^{-1/2}\;;
\end{equation}
in general
\begin{equation}
  \expcth{|W(p,q)|^n} \approx \frac{2^{-n/2} n!}{\Gammalr{\frac n 2 + 1}} d^{-3n/2}\;.
\end{equation}
\eqref{eq:pure-first-moment} gives Wigner norm
\begin{equation}
  \label{eq:pure-wigner-norm}
  \expcth{\mathcal W} = \frac{d!!}{(d-1)!!}\;.
\end{equation}
We see in Figs.~\ref{fig:pure-wigner-norm} and \ref{fig:pure-mana} that this precisely matches the numerics.

\section{Distinguishing the Haar distribution from an approximate $t$-design} \label{s:tdesign}

The Haar distribution can be approximated well by the output of random circuits with very few magic states.
Such an approximation is called a \textbf{$t$-design}.
A $t$-design is a set $K$ of unitaries such that
for any polynomial $P_{(t,t)}(U)$ of degree at most $t$ in the matrix elements $U_{ij}$ and their conjugates $U_{ij}^*$,
\begin{equation}
  \sum_{U \in K} P_{(t,t)}(U) = \int d\mu(U) P_{(t,t)}(U)\;
\end{equation}
where $\mu(U)$ is the Haar measure on unitaries \cite{dankert_exact_2009}.
That is, the uniform distribution over $K$ matches the first $t$ moments of the Haar distribution.
Clifford circuits form 2-designs for general qudits and 3-designs for qubits \cite{webb_clifford_2016,zhu_multiqubit_2017}.
For multi-qubit systems
approximate $t$-designs with error $\epsilon$ have been constructed
using
\begin{equation}
  \label{eq:approx-t-design}
  \text{\# non-Clifford} = O(t^4\lg t \lg(1/\varepsilon))
\end{equation}
non-Clifford gates, independent of the number of qubits \cite{haferkamp_quantum_2020}.
The authors of that work believe that their construction will generalize to qudits.

Our result---that most states chosen from a Haar distribution have nearly maximal magic---is therefore surprising.
One might expect properties of the Haar distribution
to broadly reflect those of the Clifford unitary 2-design
(or the approximate $t$-designs);
since those states clearly have low mana,
this would lead one to expect most Haar states to have very little mana.
We find the opposite.

In fact, if \eqref{eq:approx-t-design} holds for qudits,
then for sufficiently large $d$
one can distinguish a Haar distribution
from the resulting approximate $t$-design with high probability
by sampling just one unitary, applying it to a computational basis state, measuring the Wigner norm of the result.
Recall that the overwhelming majority of states taken from the Haar distribution have Wigner norm close to
\begin{equation}
  \mathcal W = \frac{d!!}{(d-1)!!} \approx \sqrt{2\pi}\; d^{1/2}\;.
\end{equation}
To quantify that statement,
take the Gaussian estimate \eqref{eq:g-var}
\begin{equation}
  \expcth{\Delta \mathcal W^2} = d^{-1}(1 - 2/\pi)
\end{equation}
to be an upper bound on the variance of the Wigner norm in the Haar distribution
and apply Chebyshev's (second) inequality.
Then the probability that a Haar state has Wigner norm less than
\begin{equation}
  \expcth{\mathcal W} - \delta
\end{equation}
is
\begin{equation}
  \label{eq:chebyshev-inequality}
  P\left[\mathcal W < \expcth{\mathcal W} - \delta\right] \le \frac 1 {d\delta^2}
\end{equation}
If the state is the output of the $t$-design,
by contrast,
it will have Wigner norm at most
\begin{equation}
  \mathcal W_{\text{$t$-design}} = \varepsilon^{-ct^4 \lg t}
\end{equation}
for some constant $c$.
The probability that one might confuse a Haar state for the output of a $t$-design is therefore
\begin{align}
  P\left[\mathcal W < \mathcal W_{\text{$t$-design}}\right]
  &\le \left(d \left[\sqrt{2/\pi}\;d^{1/2} - \varepsilon^{-ct^4 \lg t}\right]\right)^{-1}\notag \\
  &\sim d^{-3/2}
  \label{eq:p-distinguish}
\end{align}
for $d \gg \varepsilon^{-2ct^4\lg t}$.
We have freely used the asymptotic approximation $d!! / (d-1)!! \approx \sqrt{2/\pi}\; d^{1/2}$ for the Haar average.
This is analogous to the result
\begin{equation*}
  \label{eq:liu-winter}
  P(\mathfrak D_{\min}(\psi) < n - 2\log n - 0.63) < e^{-n^2}
\end{equation*}
of \cite{liu_many-body_2020} for
the min-relative entropy of magic $\mathfrak D_{\min}(\psi)$
of random pure state $\psi$
on an $n$-qubit subsystem.
(We expect their result to extend to qudits, modulo a change of constants.)
To roughly translate between their result \eqref{eq:liu-winter} and ours \eqref{eq:p-distinguish}, take
$n \sim \lg d$.

How can our result \eqref{eq:p-distinguish} be?
One might expect the fact that the mana distinguishes the Haar distribution from an approximate $t$-design
to result from some subtle concentration of measure argument,
because the probability calculation \eqref{eq:p-distinguish}
rests on the Chebyshev's inequality result \eqref{eq:chebyshev-inequality},
which could be Levy's lemma \cite{milman_asymptotic_2001} in disguise.
But this is not the case:
the Wigner norm varies too widely for a straightforward application of Levy's lemma to give a result like \eqref{eq:chebyshev-inequality}.
(We discuss this in App.~\ref{sec:conc-meas-vari}.)

Mana distinguishes between the two distributions
because the cusp in the absolute value function
means that the expansion of the Wigner norm in Chebyshev polynomials---which is morally the Fourier series---of the Wigner function has large weight on high-order polynomials.
% since the Wigner function is a second-order polynomial in the state,
% this means that the Wigner norm has large weight on high-order polynomials of the state,
% and hence the unitary that constructs the state.
To be more precise,
\begin{equation}
  \mathcal W = \sum_{pq} \sum_n \tau_n T_n[W(p,q)]
\end{equation}
with $T_n(x) = \cos(n\cos^{-1}(x))$ the Chebyshev polynomials.
The Chebyshev expansion is legitimate because the purity constraint \eqref{eq:wigner-sum-sq} give $-1 \le W(p,q) \le 1$.
Then the $\tau_n$ are precisely the Fourier coefficients in the expansion
\begin{equation}
  |\cos\theta| = \tau_0 + \frac 1 2 \sum_{n\ne 0} \tau_n e^{in\theta}\;,
\end{equation}
i.e.\
\begin{align}
  \label{eq:W-chebyshev-coeff}
  \begin{split}
    \tau_{2k} &= \frac 8 \pi \frac {(-1)^k} {(2k)^2 - 1}\\
    \tau_{2k + 1} &= 0\;.
  \end{split}
\end{align}
Explicitly
\begin{align}
  \mathcal W = \sum_k
  &\frac 8 \pi
  \frac {(-1)^k} {(2k)^2 - 1}\\
  &\times\sum_{pq} T_{2k}\left[d^{-1}\sum_x\omega^{-px} \psi_{q + 2^{-1} x} \psi^*_{q - 2^{-1} x}\right]\notag\;:
\end{align}
that is, $\mathcal W$ has weight $\sim n^{-2}$ on $n$th-order polynomials.

\section{Discussion}\label{s:discussion}
We have computed the average Wigner norm of random states, pure and mixed;
in so doing we have quantified the amount of non-Clifford resources required to construct random states.
We found that the Wigner norm is controlled by  the entropy deficit
\[ \Delta := \ell \ln d - S_2 \;, \]
where $\ell$ is the number of qudits,
$d$ the qudit dimension,
and $S_2$ the state's second \Renyi entropy.
We further found that the almost all states have Wigner norm very close to the average,
so we can take the mana to be
\[\expcth{\mathcal M} \equiv \expcth{\ln \mathcal W} \approx \ln \expcth{W}\;; \]
for states not close to being maximally mixed this very well approximated by
\[\expcth{\mathcal M} = \frac 1 2 [\Delta - \ln \pi / 2]\;. \]
This lays the groundwork for further calculations of the mana of Haar states
and serves as a benchmark for studies of magic in thermalization, black holes, and other physical systems.

Future directions include the study of the analytical properties of magic monotones.
Our work and \cite{liu_many-body_2020}, which was posted as we finished this work, both show
that most Haar states on $n$ qudits
require an extensive number of non-Clifford gates to prepare.
But there exist approximate $t$-designs requiring a constant-in-size number of non-Clifford gates \cite{haferkamp_quantum_2020}.
Any good magic monotone should distinguish these two situations,
and be sensitive to arbitrarily high moments of the distribution of wavefunctions.
What analytic properties should good magic monotones have?
Conversely,
can we use the fact that
approximate $t$-designs (approximately) share the first $t$ moments of the Haar distribution,
together with the characteristic functions \eqref{eq:mixed-characteristic-exact}, \eqref{eq:saddle-point-char}
to lower-bound the number of non-Clifford gates required to create an approximate $t$-design? 

Our work also raises larger questions about the relationship of magic with correlations and entropies, and entanglement.
Consider a random pure state $\psi$
on the tensor product of two systems $A,B$
with Hilbert space dimensions $d_A = d_B = d \gg 1$.
The mana of the whole system will be
\[ \mathcal M_{AB} = \ln d - \frac 1 2 \ln \pi / 2\;, \]
but the mana of each subsystem will be very roughly
\[ \mathcal M_A = \mathcal M_B \approx \ln 2/\sqrt{\pi} \ll \frac {\mathcal M_{AB}} 2 \]
since $A$ and $B$ will each have entropy $S_2 \approx \ln d - \ln 2$. 
Mana is therefore best thought of as residing in the correlations between $A$ and $B$.
But how can we make this notion more precise?
How does it constrain (for example) tensor network approximations to physical states resembling Haar states?
Can we distill this magic (cf.\ \cite{cao_inprog})?
These questions were already implicit in \cite{white_conformal_2020},
but here we see them stripped of extraneous effects.

\begin{acknowledgements}

We are grateful to Zak Webb, for useful discussion of unitary $k$-designs,
Mike Winer, for helpful conversations about the Gaussian approximation,
and Charles Cao and Brian Swingle for many stimulating conversations about magic.

CDW gratefully acknowledges funding from the U.S. Department of Energy (DOE), Office of Science, Office of Advanced Scientific Computing Research (ASCR) Quantum Computing Application Teams program, for support under fieldwork proposal number ERKJ347.
  
\end{acknowledgements}

\bibliography{references}
\appendix

\section{Generalized Pauli and phase space point operators}\label{app:pauli-phase}
\subsection{Generalized Pauli operators}

Define \textbf{shift} and \textbf{clock} operators
\begin{align*}
X &= \ketbra{j + 1 \mod d}{j} \\
Z &= \omega^j \ketbra{j}{j}\; ,
\end{align*}
$\omega = e^{2\pi i / d} $.
These operators commute like
\begin{equation*}
ZX = \omega XZ
\end{equation*}
for
\begin{equation*}
Z^{a_1} X^{a_2} = \omega^{a_1a_2} X^{a_2} Z^{a_1}\;.
\end{equation*}
Then the \textbf{generalized Pauli} or \textbf{Weyl} operators are
\begin{align*}
T_{a_1, a_2} =  \omega^{-2^{-1}a_1a_2} Z^{a_1} X^{a_2} 
\end{align*}
with $a_1, a_2 \in \mathbb Z^d$ and $2^{-1}$ the inverse of $2$ in the field $\mathbb Z^d$
\begin{equation}
  2^{-1} = \frac {d+1} 2\;.
\end{equation}
We will use notation $\bm a = (a_1, a_2)$ etc. in e.g. $T_{\bm a} \equiv T_{a_1 a_2}$.

\subsection{Phase space point operators}

The \textbf{phase space point operators} are the operators $A_{\bm u} \equiv A_{u_1 u_2}$ such that 
\begin{equation}
  W(u_1,u_2) = \bra \psi A_{u_1, u_2} \ket \psi\;.
\end{equation}
They can be written
\begin{subequations}
  \begin{align}
    A_{\bm 0} &= d^{-1}\sum_{\bm a} T_{\bm a}\\
    A_{\bm u} &= T_{\bm u} A_{\bm 0} T_{\bm u}^\dagger \;.
  \end{align}
\end{subequations}
The phase space point operators are Hermitian and unitarily equivalent.

To find the spectrum of $A_{pq}$, permute the basis by
\begin{align}
  \begin{split}
    q &\mapsto 0\\
    q + x &\mapsto 2x - 1\\
    q - x & \mapsto 2x\;,
  \end{split}
\end{align}
$x > 0$.
In this basis
\begin{equation}
  d A = 1 \oplus \bigoplus_{x = 1}^{(d-1)/2} \begin{bmatrix} 0 & \omega^{px} \\ \omega^{-px} & 0 \end{bmatrix}\;.
\end{equation}
Immediately the spectrum of $A$ is
\begin{equation}
  \mathrm{spec}\;A(p,q) =
  \{
  \underbrace{d^{-1}, \dots, d^{-1}}_{\text{$(d + 1)/2$} },
  \underbrace{-d^{-1}, \dots, -d^{-1}}_{\text{$(d - 1)/2$} }
  \}\;.
\end{equation}

\section{Concentration of measure and variability of mana}\label{sec:conc-meas-vari}
In Sec.~\ref{s:tdesign}
we argued that the mana of a Haar random state has very little variability when the Hilbert space dimension is large---that is, that almost all states have Wigner norm very near the mean.
One might expect this to result merely from large-dimensional concentration of measure.
In fact it does not:
rather, it is a result of the cusp in the Wigner norm.
In this appendix we give the putative concentrarion-of-measure argument, and show how it fails.

For large Hilbert space dimension,
one expects the overwhelming majority of states will have almost the same value for any reasonable function $f$.
(We follow \cite{milman_asymptotic_2001}; we again drop the bra-ket notation for states.) 
To be more precise, first let $f : \mathds CP^{d} \rightarrow \mathds R$ be continuous.
Write $M_f$ for its median
\begin{align}
  &\mu(\{\psi : f(\psi) \le M_f\}) \ge 1/2\ \text{and}\\
  &\mu(\{\psi : f(\psi) \ge M_f\}) \ge 1/2
\end{align}
and
\begin{align}
  A = \{\psi: f(\psi) = M_f\}
\end{align}
for the states that have the median value.
Then \textit{Levy's lemma} states that for large $d$ the overwhelming majority of states are close to $A$,
in the sense that
\begin{align}
  \mu(\{\psi: d(\psi, A) \le \epsilon\}) \ge 1 - \sqrt{\pi/2}\; e^{-\epsilon^2 d}
\end{align}
where $d(\cdot, \cdot)$ is the geodesic metric.
If in addition $f$ is Lipshitz-continuous with Lipshitz constant $K$,
that is
\begin{equation}
  |f(\psi) - f(\phi)| \le K d(\psi,  \phi),
\end{equation}
then the overwhelming majority of states $\psi$ have $f(\psi)$ close to $M_f$ in the sense that
\begin{align}
  \begin{split}
    \mu(\{\psi : |f(\psi) - M_f| < \epsilon\} &\ge \mu(\{\psi : d(\psi, A) < K^{-1}\epsilon\})\\
    &\ge 1 - \sqrt{\pi/2} e^{-(K^{-1}\epsilon)^2 d}\;.
    \end{split}
\end{align}
By a non-rigorous physicists' estimate, then, we can integrate with 
\begin{equation}
  d\mu(\epsilon) \le \sqrt{\pi/2}\; dK^{-2}\epsilon e^{-K^{-2}d\epsilon^2} d\epsilon\;.
\end{equation}
This means that the mean is asymptotically close to the median
\begin{align}
  \begin{split}
    |\expcth{f - M_f}|
    &\equiv \left| \int_0^\infty d\mu(\epsilon)\; \epsilon \right|\\
    &\le  \sqrt{\pi/2}\; dK^{-2} \int_0^\infty d\epsilon\;\epsilon^2 e^{-dK^{-2} \epsilon^2}\\
    &= \frac{\pi K}{\sqrt{8d} }
\end{split}
\end{align}
and the variance is
\begin{align}
  \begin{split}
  \expcth {f^2} - \expcth{f}^2
  &\le \expcth{(f - M_f)^2}\\
  &= \int_0^\infty d\mu(\epsilon) \epsilon^2\\
  &\le \sqrt{\pi/2} dK^{-2} \int_0^\infty d\epsilon \epsilon^3 e^{-dK^{-2} \epsilon^2}\\
  &\propto K^2/d\;.
\end{split}
\end{align}

To apply these estimates to the Wigner norm and the mana we need the Lipshitz constant $K$ for each.
Consider first the Wigner norm.
Write $\psi_{\max}$ for the state with maximum Wigner norm.
Since the average Wigner norm has the form $\expcth{\mathcal W} = d^{1/2} - \gamma$ for some $\gamma$,
we can take $\|\psi_{\max}\|_W >  d^{1/2} - \gamma$.
But because the maximum geodesic distance between two points is $\pi$, $\psi_{\max}$ is at most a distance $\pi$ away from a stabilizer state.
So the Lipshitz constant for the Wigner norm is at least
\begin{align}
  K \ge \pi^{-1}(d^{1/2} - \gamma) \sim d^{1/2}
\end{align}
and the key parameter in the concentration-of-measure results is
\begin{align}
  K^{-2} d \sim const.
\end{align}
We can therefore cannot appeal to concentration of measure: we need extra information about the Wigner norm.
That extra information is that it is zero only for stabilizer states (that is, on a set of measure zero).
Basically the Wigner function itself is almost constant through most of the sphere.

\end{document}